\documentclass[preprint,preprintnumbers,amsmath,amssymb]{revtex4}

\usepackage{epsfig}
\usepackage{amssymb, amsmath}
\usepackage{color}
\usepackage{ulem}
\def\rev#1{\textcolor{black}{#1}}

\begin{document}

\title{Marchenko-based target replacement, accounting for all orders of multiple reflections}

\author{Kees Wapenaar and Myrna Staring}
\affiliation{Department of Geoscience and Engineering, Delft University of Technology, P.O. Box 5048, 2600 GA Delft, The Netherlands}


\def\bx{{\bf x}}
\def\bxr{{\bf x}_R}
\def\bxs{{\bf x}_S}
\def\bxh{{\bf x}_{\rm H}}
\def\bxhr{{\bf x}_{{\rm H},R}}
\def\bxhs{{\bf x}_{{\rm H},S}}
\def\setdD{\mathbb{S}}
\def\setR{\mathbb{R}}
\def\half{\begin{matrix}\frac{1}{2}\end{matrix}}
\def\pa {{\bf p}_A}
\def\pb {{\bf p}_B}
\def\f {{\bf F}_{1,A}}
\def\ff {{\bf F}_{2,A}}
\def\fb {{\bf F}_{1,B}}
\def\ffb {{\bf F}_{2,B}}
\def\R {{\bf R}}
\def\T {{\bf T}}
\def\G {{\bf G}}
\def\M {{\bf C}}
\def\I {{\bf I}}
\def\O {{\bf O}}
\def\r{{\bf r}^\cap}
\def\too{\to}

\def\Rbb{{\bar\R}_b}
\def\RBb{{\bar\R}_B}
\def\RCb{{\bar\R}_C}

\def\Rb{{\bar R}_b}
\def\RB{{\bar R}_B}
\def\RC{{\bar R}_C}

\def\Tbb{{\bar\T}_b}
\def\TBb{{\bar\T}_B}

\def\GBb{{\bar\G}_B}
\def\GCb{{\bar\G}_C}

\def\MAbb{{\bar\M}_{Ab}}
\def\MBcb{{\bar\M}_{Bc}}

\begin{abstract}
In seismic monitoring\rev{,} one is usually  interested in the response of a changing target zone, embedded in a static inhomogeneous medium. 
We introduce an efficient method which predicts reflection responses at the earth's surface for different target-zone scenarios, 
 from a single reflection response at the surface and a model of the changing target zone. 
The proposed process consists of two main steps. 
In the first step, the response of the original target zone is removed from the reflection response, using the Marchenko method. In the second step, the modelled response of a new target zone 
is inserted between the overburden and underburden responses. The method fully accounts for all orders of multiple scattering and, in the elastodynamic case, for wave conversion.
For monitoring purposes, only the second step needs to be repeated for each target-zone model. 
Since the target zone covers only a small part of the entire medium, the proposed method is much more efficient than repeated modelling of the  entire reflection response.
\end{abstract}

\maketitle

\section{Introduction}

In seismic modelling, inversion\rev{,} and monitoring one is often interested in the response of a relatively small target zone, embedded in a larger inhomogeneous medium. 
Yet, to obtain the seismic response of a target zone at the earth's surface, the entire medium enclosing the target should be involved in the modelling process.
This may become very inefficient when different scenarios for the target zone need to be evaluated, or when a target that changes over time needs to be monitored, for example to follow
 fluid flow in an aquifer, subsurface storage of waste products\rev{,} or production of a hydrocarbon reservoir.
Through the years, several efficient methods have been developed for modelling successive responses of a medium in which the parameters change only in a target zone.
\citet{Robertsson2000GEO} address this problem with the following approach. First they model the wave field in the full medium,
define a boundary around the target zone in which the changes take place, and evaluate the field at this boundary.
Next, they numerically inject this field from the same boundary into different models of the target zone.
Because the target zone usually covers only a small part of the full medium, 
this injection process takes only a fraction of the time that would be needed to model the field in the full medium. 
This method is very well suited to model different time-lapse scenarios of a specific subsurface process in an efficient way. 
A limitation of the method is that multiple scattering between the changed target and the embedding medium is not taken into account.
The method was adapted by \citet{VanManen2007JASA} 
to account for this type of interaction, by modifying the field at the boundary around the changed target at every time-step of the simulation.
Wave field injection methods are not only useful for efficient numerical modelling of wave fields in a changing target zone, 
they can also be used to physically inject a field from a large numerical environment into a finite-size physical model \citep{Vasmel2013JASA}. 

Instead of numerically modelling the field at the boundary enclosing the target, 
\citet{Elison2016EAGE} propose to use the Marchenko method to derive this field from reflection data
at the surface. Hence, to obtain the wave field in a changing target zone, they need a measured reflection response at the surface of the original medium and a model of the target.
Their method exploits an attractive property of the Marchenko method, namely that ``redatumed'' reflection responses of a target zone from above ($\R^\cup$) and from below ($\R^\cap$)
can both be obtained from single-sided reflection data at the surface and an estimate of the direct arrivals between the surface and the target zone  \citep{Wapenaar2014GEO}.

In most of the methods discussed above, the wave fields are derived inside the changing target.
Here we discuss a method which predicts  reflection responses (including all multiples) at the earth's surface for different target-zone scenarios,   
from a single reflection response at the surface 
and a model of the changing target zone. 
The proposed method, which we call ``target replacement'', consists of two main steps. In the first step, which is analogous to the method proposed by \citet{Elison2016EAGE}, 
we use the Marchenko method to remove the response of the target zone from the original reflection response.  
In the second step we insert the response of a new target zone, yielding the desired reflection response at the surface for the particular target-zone scenario.
Both steps fully account for multiple scattering between the target and the embedding medium. 
Note that, to model different reflection responses for different target models, only the second step needs to be repeated. 
Hence, this process is particularly efficient when reflection responses at the surface are needed for many target-zone scenarios.
\rev{Also note that, unlike the model-driven methods of \citet{Robertsson2000GEO} and \citet{VanManen2007JASA}, our method as well as that of \citet{Elison2016EAGE} 
only needs a smooth model of the overburden and no model of the underburden. The required detailed information of the over- and underburden comes from the measured reflection response.}

\rev{Similar as the other methods discussed in this introduction, we assume that the target zone is the only region in which changes occur; 
the over- and underburden are assumed to remain unchanged. However,  changes in a reservoir may 
lead to changes in the embedding medium \citep{Hatchell2005TLE, Herwanger2009GEO}. When this is the case, the target zone should not be restricted to the reservoir, but it should also include
the part of the embedding medium in which the changes  have a noticeable effect on the waves propagating through it. 
Of course the larger the target zone, the smaller the efficiency gain.}

The setup of this paper is as follows. In \rev{S}ection \ref{sec2}\rev{,} we derive a representation of the seismic reflection response at the earth's surface (including all orders of multiple scattering), 
which explicitly distinguishes between the response of the target zone and that of the embedding medium. 
Next, based on this representation, in \rev{S}ection \ref{sec3}\rev{,} we discuss how to remove the response of the target zone from the reflection response at the surface.
 In \rev{S}ection \ref{sec4}\rev{,} we discuss how the response of a changed target zone can be inserted into the reflection response at the surface.
 The proposed method is illustrated with numerical examples in \rev{S}ection \ref{sec5}. \rev{We end the paper with a discussion  (\rev{S}ection \ref{sec6}) and conclusions (\rev{S}ection \ref{sec7})}. 

\section{Representation of the reflection response}\label{sec2}

 \begin{figure}
\centerline{\epsfxsize=10 cm\epsfbox{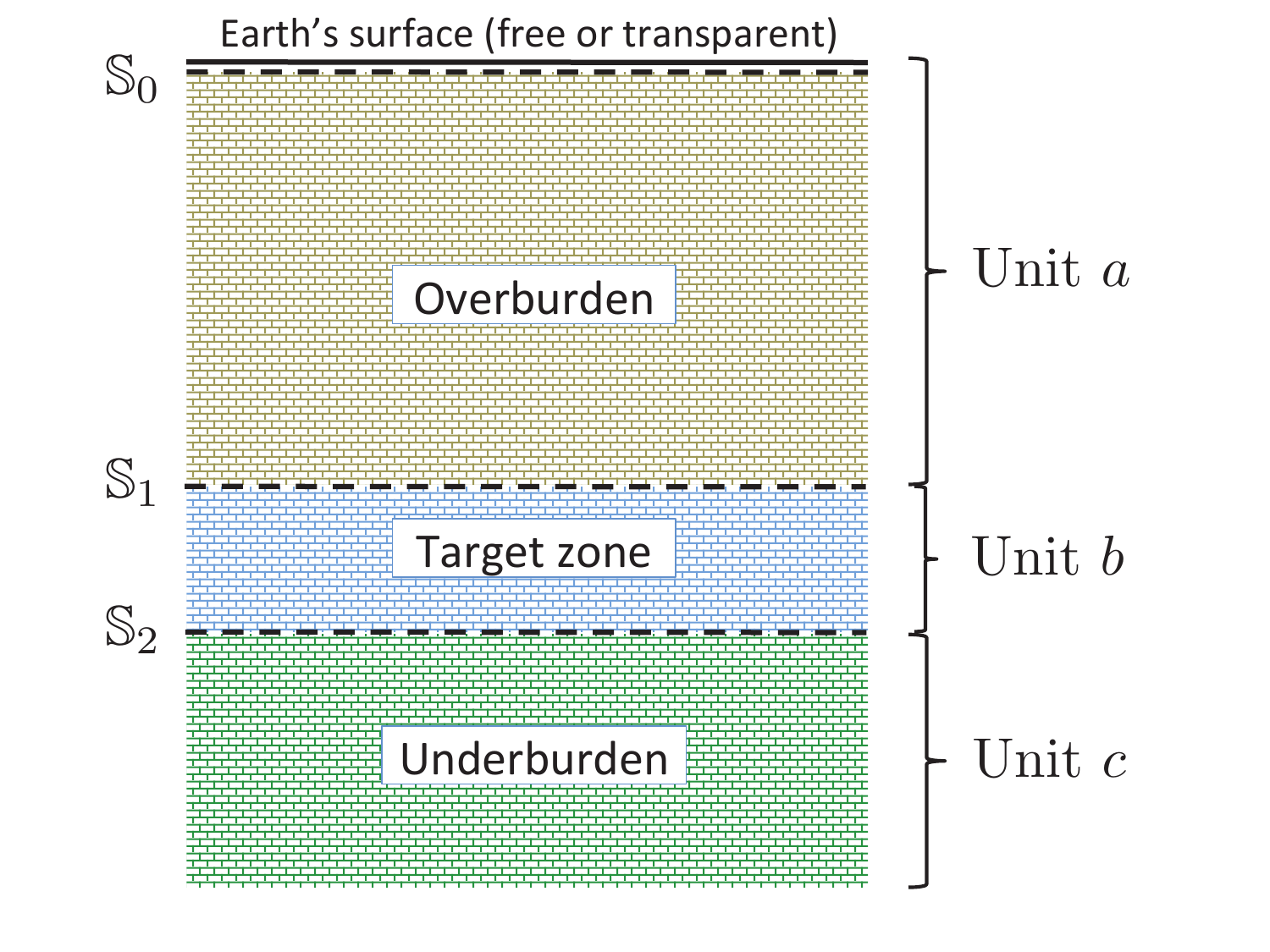}}
\vspace{-.3cm}
\caption{
Subdivision of the inhomogeneous subsurface into three units: an overburden (unit $a$), a target zone (unit $b$) and an underburden (unit $c$). 
Note that unit $a$ includes the earth's surface just above $\setdD_0$. This surface may be considered either as a free or as 
a transparent surface.}\label{Fig1}
\end{figure}

We derive a representation for the reflection response at the earth's surface, which distinguishes between the response of the target zone and that of the embedding medium.
We start by dividing the subsurface into three units. The first unit, indicated as unit $a$ in Figure \ref{Fig1}, covers the region between the earth's surface and boundary $\setdD_1$, 
the latter defining the upper boundary of the target zone.
The earth's surface (indicated by the solid line) may be considered either as a free or as a transparent surface (the latter after surface-related multiple elimination). 
The earth's surface is included in unit $a$. A transparent boundary 
$\setdD_0$ (indicated by the upper dashed line) is defined at an infinitesimal distance below the earth's surface  
(in the following we abbreviate ``an infinitesimal distance above/below'' as ``just above/below'').
Unit $a$, i.e.,  the region above the target zone, is called the overburden. 
The second unit,  indicated as unit $b$ in Figure \ref{Fig1}, represents the target zone and is enclosed by boundaries $\setdD_1$ and $\setdD_2$. The third unit,
indicated as unit $c$ in Figure \ref{Fig1}, represents the region below the lower boundary of the target zone, $\setdD_2$. Unit $c$, i.e., the region below the target zone, is called the underburden.

We assume that the media inside the units are arbitrary inhomogeneous, lossless media. 
Furthermore, we assume that the boundaries  $\setdD_1$ and $\setdD_2$ do not coincide with interfaces,
or in other words, we consider these boundaries to be transparent for downgoing and upgoing waves incident to these boundaries. 
The representation derived below could be extended to account for scattering 
at these boundaries, but that would go at the cost of clarity. By allowing some flexibility in the definition of the target zone, 
it will often be possible to choose boundaries $\setdD_1$ and $\setdD_2$ that are (close to) transparent.

The starting point for the derivation of the representation and the target replacement method is formed by the following one-way reciprocity theorems in the space-frequency domain
\begin{equation}\label{eq1}
\int_{\setdD_m}\{(\pa^+)^t \pb^- - (\pa^-)^t\pb^+\}{\rm d}\bx = \int_{\setdD_n}\{(\pa^+)^t \pb^- - (\pa^-)^t\pb^+\}{\rm d}\bx
\end{equation}
and
\begin{equation}\label{eq1b}
\int_{\setdD_m}\{(\pa^+ )^\dagger\pb^+ - (\pa^-)^\dagger\pb^-\}{\rm d}\bx = \int_{\setdD_n}\{(\pa^+ )^\dagger\pb^+ - (\pa^-)^\dagger\pb^-\}{\rm d}\bx
\end{equation}
\citep{Wapenaar96GJI1}. Here $\setdD_m$ and $\setdD_n$ can stand for any of the boundaries $\setdD_0$, $\setdD_1$ and $\setdD_2$.
Subscripts $A$ and $B$ refer to two independent states.   
Superscripts $+$ and $-$ stand for downward  and upward propagation, respectively. 
Superscript $t$ in equation (\ref{eq1}) denotes \rev{the transpose} and superscript $\dagger$ in equation (\ref{eq1b}) \rev{the adjoint (i.e., the complex conjugate transpose)}.
The vectors $\pa^\pm$ and $\pb^\pm$ represent flux-normalised one-way wave fields in states $A$ and $B$. For the elastodynamic situation they are defined as
\begin{equation}\label{eqp}
\pa^\pm(\bx,\omega)=\begin{pmatrix} \Phi_A^\pm \\ \Psi_A^\pm \\ \Upsilon_A^\pm \end{pmatrix}\!(\bx,\omega),\quad
\pb^\pm(\bx,\omega)=\begin{pmatrix} \Phi_B^\pm \\ \Psi_B^\pm \\ \Upsilon_B^\pm \end{pmatrix}\!(\bx,\omega),
\end{equation}
where $\Phi_{A,B}^\pm$, $\Psi_{A,B}^\pm$ and $\Upsilon_{A,B}^\pm$ represent $P$, $S1$ and $S2$ waves, respectively. For the acoustic situation\rev{,} 
$\pa^\pm(\bx,\omega)$ and $\pb^\pm(\bx,\omega)$ reduce to scalar functions.
The Cartesian coordinate vector $\bx$ is defined as $\bx=(x_1,x_2,x_3)$ (the $x_3$-axis pointing downward) and $\omega$ denotes angular frequency. 
An underlying assumption for both reciprocity theorems is that the medium parameters in states $A$ and $B$ are identical in the domain enclosed by boundaries $\setdD_m$ and $\setdD_n$.
Outside this domain the medium parameters in state $A$ may be different from those in state $B$, a property that we will make frequently use of throughout this paper. 
Another assumption is that there are no sources between $\setdD_m$ and $\setdD_n$. 
Finally, an assumption that holds specifically for equation (\ref{eq1b}) is that evanescent waves are neglected at boundaries $\setdD_m$ and $\setdD_n$.
For a more detailed discussion of these one-way reciprocity theorems, including their extensions for the situation that the domain between $\setdD_m$ and $\setdD_n$ contains sources and the medium parameters in the two states are different in this domain, see  \citet{Wapenaar96GJI1}.

\begin{figure}
\centerline{\epsfysize=16 cm\epsfbox{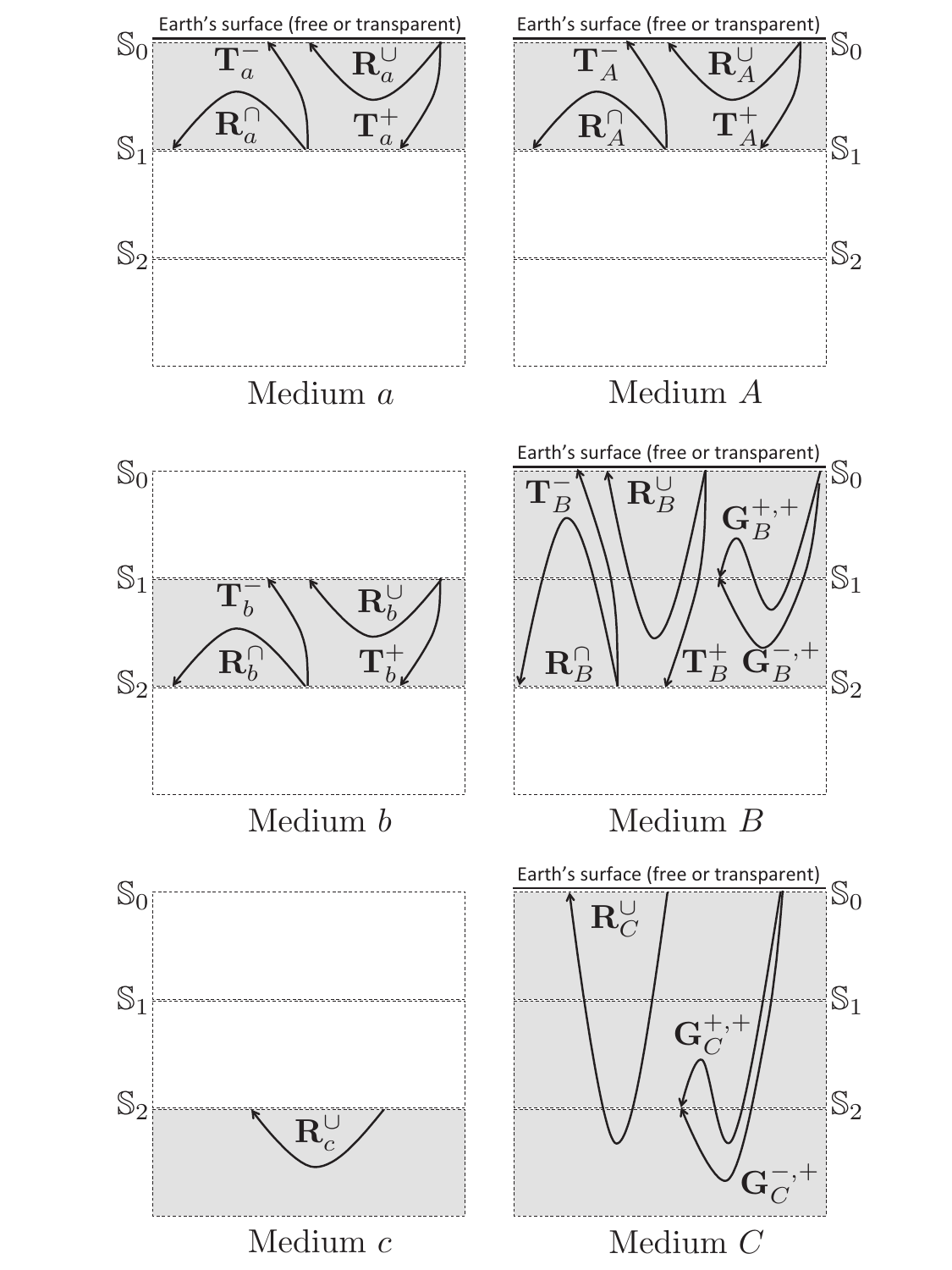}}
\caption{
Six media with their responses. Grey areas represent the inhomogeneous units (and combinations thereof) of Figure \ref{Fig1}. 
Media $A$ (=$a$), $B$ and $C$ include the earth's surface just above $\setdD_0$, which may be considered either as a free or as a transparent surface. 
The rays stand for the full responses, including all orders of multiple scattering and, in the elastodynamic case, mode conversion.}\label{Fig2}
\end{figure}

In the following derivations, equations (\ref{eq1}) and (\ref{eq1b}) will frequently be applied, each time to a combination of independent wave states in two media that are identical in the domain between
$\setdD_m$ and $\setdD_n$. 
Figure \ref{Fig2} shows six media that will be used in different combinations. 
Media $a$, $b$ and $c$ in the left column contain the units $a$ (the overburden), $b$ (the target zone) and $c$ (the underburden) of the actual medium, 
each embedded in a homogeneous background. The grey areas indicate the inhomogeneous units (as depicted in Figure \ref{Fig1}), whereas the white areas represent the homogeneous embedding. Reflection and transmission responses are also indicated in Figure \ref{Fig2}.
Reflection responses from above and from below are denoted by $\R^\cup$ and $\R^\cap$, respectively, and the transmission responses by $\T^+$ and $\T^-$. 
 The subscripts $a$, $b$ and $c$ refer to the units to which these responses belong. The rays are simplifications of the actual responses, which contain 
 all orders of multiple scattering and, in the elastodynamic case, mode conversion. 
 When the earth's surface  just above $\setdD_0$ is a free surface, then the responses in unit $a$ also include multiple scattering related to the free surface. 
 Media $A$, $B$ and $C$ in the right column in Figure \ref{Fig2} consist of one to three units, as indicated 
 (note that medium $A$ is identical to medium $a$, whereas medium $C$ represents the entire medium).
 The reflection and transmission responses of these media 
 are indicated by capital subscripts $A$, $B$ and $C$. In addition, the Green's functions $\G^{+,+}$ and $\G^{-,+}$ in these media
 between $\setdD_0$ and the top boundary of the deepest unit are shown (the superscripts will be explained later). Again, all responses contain 
 all orders of multiple scattering (and mode conversion), including surface-related multiples when there is a free surface just above $\setdD_0$.

\begin{table}
\caption{Quantities to derive a representation for $\R_B^\cup$.}\label{table1}
\begin{center}
\begin{tabular}{ccc}
\hline
& State $A$: & State $B$: \\
&Medium $A$&Medium $B$\\
&Source at $\bxr$ just above $\setdD_0$&Source at $\bxs$ just above $\setdD_0$\\
\hline
 $\setdD_0$  & $\pa^+(\bx,\omega)\too\I\delta(\bxh-\bxhr)$ & $\pb^+(\bx,\omega)\too\I\delta(\bxh-\bxhs)$\\
 &$\hspace{1.2cm}+\r\R_A^\cup(\bx,\bxr,\omega)$&$\hspace{1.2cm}+\r\R_B^\cup(\bx,\bxs,\omega)$\\
  & $\pa^-(\bx,\omega)\too\R_A^\cup(\bx,\bxr,\omega)$ & $\pb^-(\bx,\omega)\too\R_B^\cup(\bx,\bxs,\omega)$\\
\hline
  $\setdD_1$  & $\pa^+(\bx,\omega)\too\T_A^+(\bx,\bxr,\omega)$ &  $\pb^+(\bx,\omega)\too\G_B^{+,+}(\bx,\bxs,\omega)$\\
  & $\pa^-(\bx,\omega)\too\O$ & $\pb^-(\bx,\omega)\too\G_B^{-,+}(\bx,\bxs,\omega)$\\
\hline
\end{tabular}
\end{center}\end{table}

Our aim is to derive a representation for the reflection response of the entire medium, $\R_C^\cup$, in terms of the reflection responses of media $A$ (= $a$), $b$ and $c$.
We start by deriving a representation for $\R_B^\cup$ in terms of the reflection responses of media $A$ and $b$. To this end, we substitute the quantities of Table 1 into equation (\ref{eq1}).
Let us first discuss these quantities one by one. In state $B$, the downgoing and upgoing fields in medium $B$ for $\bx$ at $\setdD_1$ are given by
\begin{equation}\label{eqex1}
\pb^\pm(\bx,\omega)\too \G_B^{\pm,+}(\bx,\bxs,\omega).
\end{equation} 
Here $\G_B^{\pm,+}(\bx,\bxs,\omega)$ is the Green's one-way wave field matrix in medium $B$ in the space-frequency domain \citep{Wapenaar96GJI2}. 
The source is at $\bxs$, which is chosen just above $\setdD_0$. The second superscript $+$ indicates that this source is downward radiating.
The receiver is at $\bx$ at $\setdD_1$. The first superscript $\pm$ indicates the propagation direction at the receiver ($+$ for downgoing and $-$ for upgoing).
Analogous to equation (\ref{eqp}), the general Green's one-way wave field matrix can\rev{,} for the elastodynamic situation\rev{,} be written as
\begin{equation}\label{eqG}
\G^{\pm,\pm}(\bx,\bx',\omega)=\begin{pmatrix}
G_{\phi,\phi}^{\pm,\pm} & G_{\phi,\psi}^{\pm,\pm} & G_{\phi,\upsilon}^{\pm,\pm}\\
G_{\psi,\phi}^{\pm,\pm} & G_{\psi,\psi}^{\pm,\pm} & G_{\psi,\upsilon}^{\pm,\pm}\\
G_{\upsilon,\phi}^{\pm,\pm} & G_{\upsilon,\psi}^{\pm,\pm} & G_{\upsilon,\upsilon}^{\pm,\pm}
\end{pmatrix}\!(\bx,\bx',\omega).
\end{equation}
Each column corresponds to a specific type of source at $\bx'$ and each row to a specific type of receiver at $\bx$ 
(where subscripts $\phi$, $\psi$ and $\upsilon$ refer to flux-normalised $P$, $S1$ and $S2$ waves, respectively).
For the acoustic situation\rev{,} $\G^{\pm,\pm}(\bx,\bx',\omega)$ reduces to a scalar function. The following reciprocity relations hold for the general Green's matrix
\begin{eqnarray}
\hspace{1.4cm}\G^{-,+}(\bx',\bx,\omega)&=&\{\G^{-,+}(\bx,\bx',\omega)\}^t,\label{eqex2}\\
\G^{+,-}(\bx',\bx,\omega)&=&\{\G^{+,-}(\bx,\bx',\omega)\}^t,\label{eqex3}\\
\G^{-,-}(\bx',\bx,\omega)&=&-\{\G^{+,+}(\bx,\bx',\omega)\}^t,\label{eqex4}
\end{eqnarray}
\citep{Haines88GJI, Kennett90GJI, Wapenaar96GJI2}. In state $B$, the upgoing field for $\bx$ at $\setdD_0$ in Table 1 is given by
\begin{equation}\label{eqex5}
\pb^-(\bx,\omega)\too \G_B^{-,+}(\bx,\bxs,\omega)= \R_B^\cup(\bx,\bxs,\omega).
\end{equation} 
Note that $\G^{-,+}(\bx,\bx',\omega)$ represents a reflection response from above, denoted by $\R^\cup(\bx,\bx',\omega)$, 
whenever the source and receiver are situated at (or just above) the same depth level.
From equations (\ref{eqex2}) and (\ref{eqex5})\rev{,} we find
\begin{equation}\label{eqex6}
 \R^\cup(\bx',\bx,\omega)=\{\R^\cup(\bx,\bx',\omega)\}^t.
\end{equation} 
Similarly, $\G^{+,-}(\bx,\bx',\omega)$ represents a reflection response from below, denoted by $\R^\cap(\bx,\bx',\omega)$, 
whenever the source and receiver  are situated at (or just below) the same depth level.
From equations (\ref{eqex3}) and (\ref{eqex5}) we find
\begin{equation}\label{eqex7}
 \R^\cap(\bx',\bx,\omega)=\{\R^\cap(\bx,\bx',\omega)\}^t.
\end{equation} 
In state $B$, the downgoing field for $\bx$ at $\setdD_0$ in Table 1 is given by
\begin{eqnarray}\label{eqex8}
\hspace{.4cm}\pb^+(\bx,\omega)&\too& \G_B^{+,+}(\bx,\bxs,\omega)
=\I\delta(\bxh-\bxhs)+\r\R_B^\cup(\bx,\bxs,\omega).
\end{eqnarray} 
Since $\bxs$ was chosen just above $\setdD_0$, the direct contribution of the flux-normalised Green's matrix $\G_B^{+,+}(\bx,\bxs,\omega)$ consists of a spatial delta function $\delta(\bxh-\bxhs)$,
with $\bxh=(x_1, x_2)$ and $\bxhs=(x_{1,S}, x_{2,S})$, hence, the singularity occurs at the lateral position of the source. This delta function is multiplied by $\I$, which is a $3\times 3$ identity matrix 
for the elastodynamic situation, to acknowledge the
matrix character of $\G_B^{+,+}(\bx,\bxs,\omega)$, as defined in equation (\ref{eqG}). For the acoustic situation $\I=1$. 
The second term in equation (\ref{eqex8}), $\r\R_B^\cup(\bx,\bxs,\omega)$, accounts for the earth's surface just above
$\setdD_0$. Here $\r$ is the reflection operator of the earth's surface from below. 
It turns the reflection response $\R_B^\cup(\bx,\bxs,\omega)$ into a downgoing field which, according to equation (\ref{eqex8}),
is added to the direct downgoing field. When the earth's surface is transparent, 
we may simply set $\r=\O$, where $\O$ is a $3\times 3$ zero matrix for the elastodynamic situation and 
$\O=0$ for the acoustic situation.
When the earth's surface is a free surface, 
$\r$ is a pseudo-differential operator for the elastodynamic situation. We introduce its transpose, $\{\r\}^t$, and adjoint, $\{\r\}^\dagger$,  via the following integral relations
\begin{equation}\label{eqex9}
\int_{\setdD_0}\{\r{\bf f}(\bx)\}^t{\bf g}(\bx){\rm d}\bx=\int_{\setdD_0}\{{\bf f}(\bx)\}^t\{\r\}^t{\bf g}(\bx){\rm d}\bx
\end{equation}
and
\begin{equation}\label{eqex10}
\int_{\setdD_0}\{\r{\bf f}(\bx)\}^\dagger{\bf g}(\bx){\rm d}\bx=\int_{\setdD_0}\{{\bf f}(\bx)\}^\dagger\{\r\}^\dagger{\bf g}(\bx){\rm d}\bx,
\end{equation}
respectively. The following properties hold \citep{Kennett90GJI, Wapenaar2004GJI}
\begin{eqnarray}
\hspace{.2cm}\{\r\}^t&=&\r,\label{eqex11}\\
\{\r\}^\dagger\r&=&\I.\label{eqex12}
\end{eqnarray}
For the acoustic situation we simply have $\r=-1$.

In state $A$, the downgoing field in medium $A$ for $\bx$ at $\setdD_1$ in Table 1 is given by
\begin{equation}\label{eqex13}
\pa^+(\bx,\omega)\too \G_A^{+,+}(\bx,\bxr,\omega)=\T_A^+(\bx,\bxr,\omega).
\end{equation} 
This time the source is at $\bxr$, again just above $\setdD_0$.
The receiver is at $\bx$ at $\setdD_1$. Note that $\G^{+,+}(\bx,\bx',\omega)$ represents a downgoing transmission response, denoted by $\T^+(\bx,\bx',\omega)$, 
whenever the source and receiver  are situated above and below an inhomogeneous slab. 
Similarly,  $\G^{-,-}(\bx',\bx,\omega)$ represents an upgoing transmission response, denoted by $-\T^-(\bx',\bx,\omega)$ (note the minus sign), 
whenever the source and receiver  are situated below and above an inhomogeneous slab. 
From equation (\ref{eqex4})\rev{,} we find
\begin{equation}\label{eqex14}
 \T^-(\bx',\bx,\omega)=\{\T^+(\bx,\bx',\omega)\}^t.
\end{equation} 
In state $A$, the upgoing field for $\bx$ at $\setdD_1$ in Table 1 is zero because medium $A$ is homogeneous below $\setdD_1$. 
The downgoing and upgoing fields in state $A$ for $\bx$ at $\setdD_0$ are defined in a similar way as in state $B$.

\begin{figure}
\centerline{\epsfysize=7 cm\epsfbox{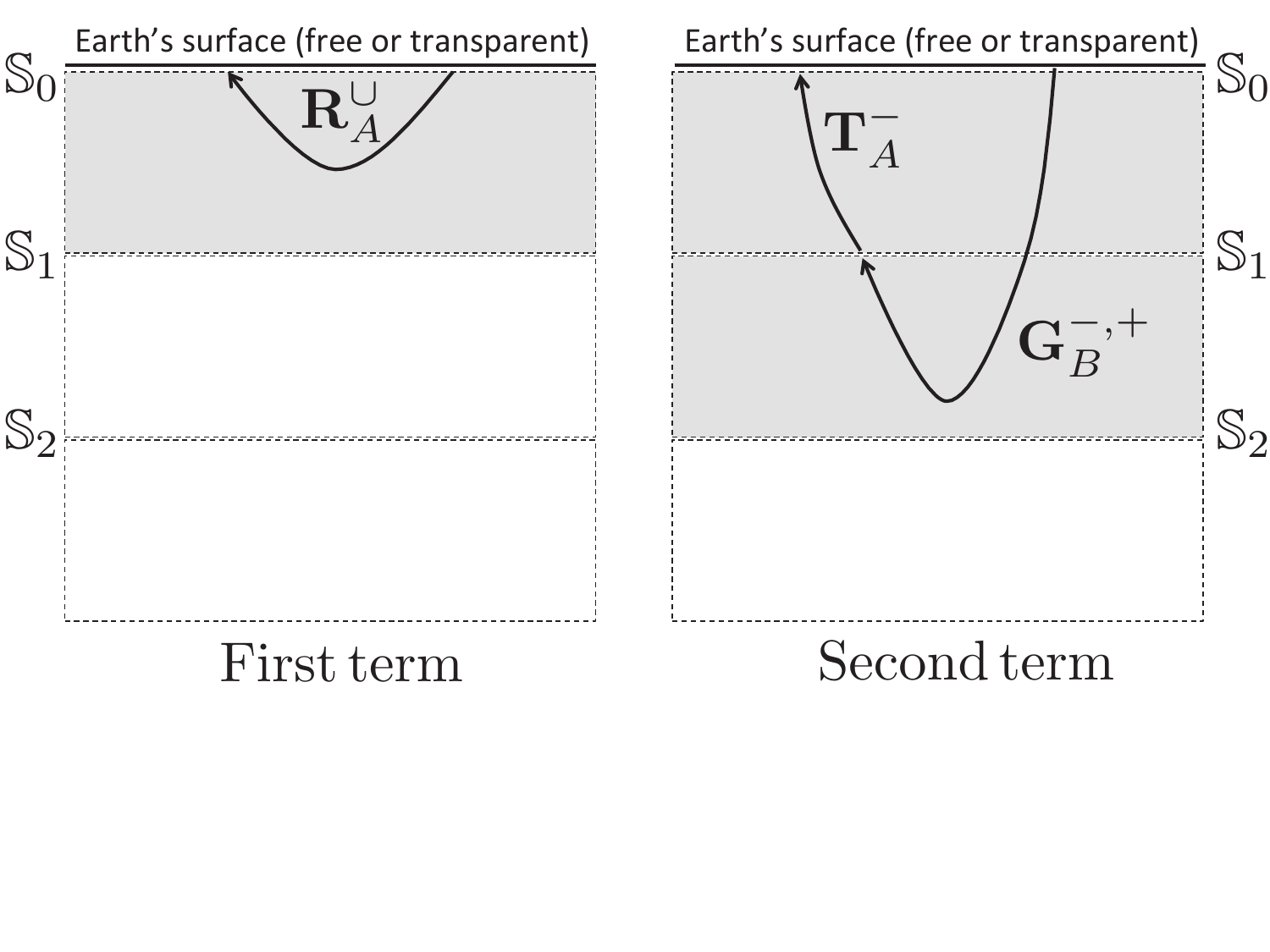}}
\vspace{-1.8cm}
\caption{
\rev{Visualization of the first an second term in the representation of equation (\ref{eq2}).}}\label{Fig3aa}
\end{figure}

Now that we have discussed all quantities in Table 1, we substitute them into equation (\ref{eq1}). Despite the different media (medium $A$ in state $A$ and medium $B$ in state $B$), this is justified, 
because between $\setdD_0$ and $\setdD_1$ these media are the same in both states (see Figure \ref{Fig2}).
Here and in the remainder of this paper, the  operator $\r$ is the same in both states (zero and thus obeying equation (\ref{eqex11}) when the earth's surface is considered transparent,
 or non-zero and obeying equations
(\ref{eqex11}) and (\ref{eqex12}) when the earth's surface is considered a free surface).
Using equations (\ref{eqex6}),  (\ref{eqex9}), (\ref{eqex11}) and (\ref{eqex14}), setting $m=0$ and $n=1$ in equation (\ref{eq1}), we obtain 
\begin{eqnarray}\label{eq2}
\R_B^\cup(\bxr,\bxs,\omega)&=&\R_A^\cup(\bxr,\bxs,\omega)
+\int_{\setdD_1}\T_A^-(\bxr,\bx,\omega)\G_B^{-,+}(\bx,\bxs,\omega){\rm d}\bx,
\end{eqnarray}
for $\bxs$ and $\bxr$ just above $\setdD_0$\rev{, see Figure \ref{Fig3aa}}.

\begin{table}
\caption{Quantities to derive a representation for $\G_B^{-,+}$.}\label{table2}
\begin{center}
\begin{tabular}{ccc}
\hline
& State $A$: & State $B$: \\
&Medium $b$&Medium $B$\\
&Source at $\bx'$ just above $\setdD_1$&Source at $\bxs$ just above $\setdD_0$\\
\hline
 $\setdD_1$  & $\pa^+(\bx,\omega)\too\I\delta(\bxh-\bxh')$ & $\pb^+(\bx,\omega)\too\G_B^{+,+}(\bx,\bxs,\omega)$\\
  & $\pa^-(\bx,\omega)\too\R_b^\cup(\bx,\bx',\omega)$ & $\pb^-(\bx,\omega)\too\G_B^{-,+}(\bx,\bxs,\omega)$\\
\hline
 $\setdD_2$  & $\pa^+(\bx,\omega)\too\T_b^+(\bx,\bx',\omega)$ &  $\pb^+(\bx,\omega)\too\T_B^+(\bx,\bxs,\omega)$\\
  & $\pa^-(\bx,\omega)\too\O$ & $\pb^-(\bx,\omega)\too\O$\\
\hline
\end{tabular}
\end{center}
\end{table}

\begin{figure}
\centerline{\epsfysize=7 cm\epsfbox{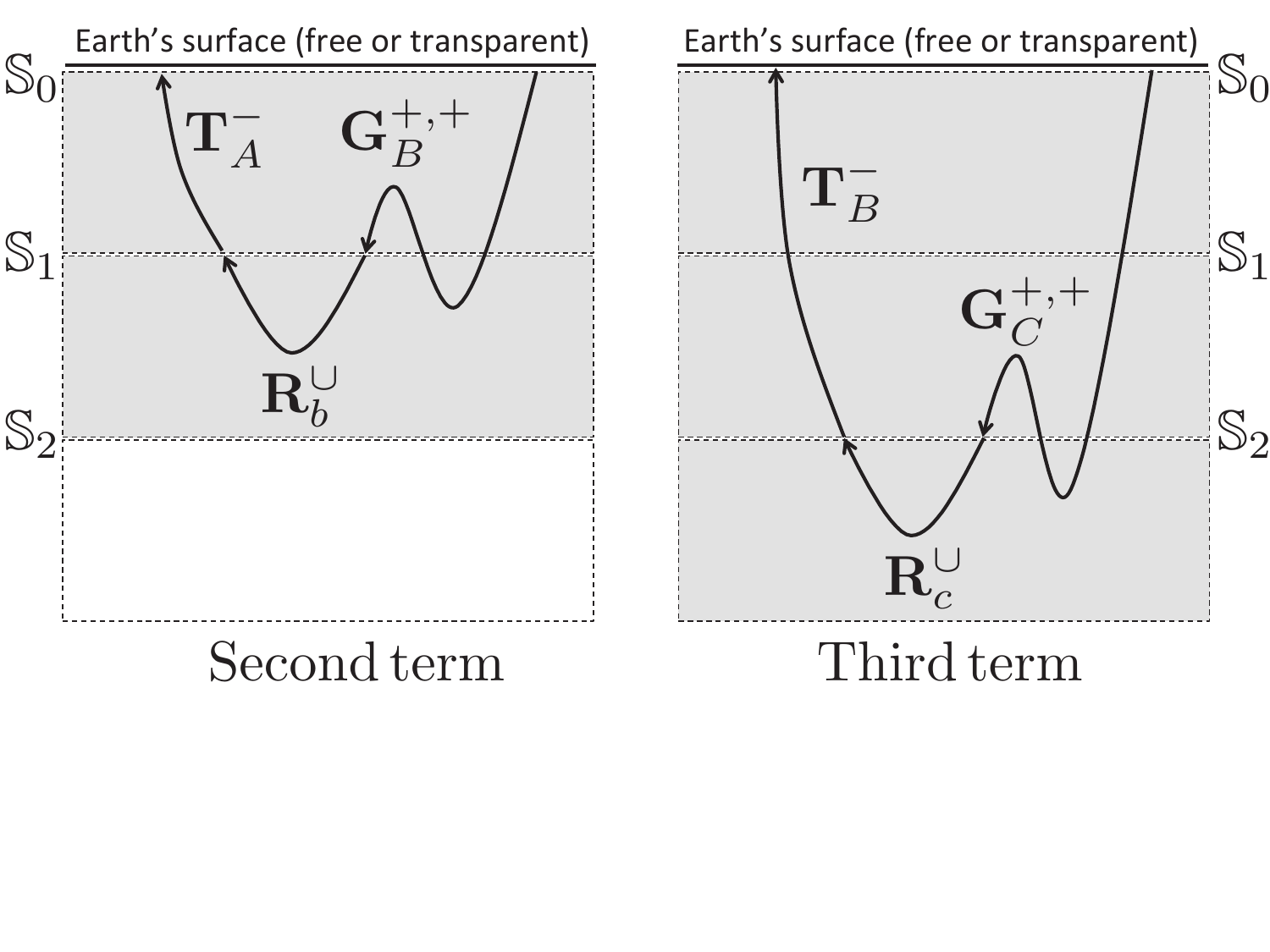}}
\vspace{-1.8cm}
\caption{
Visualization of the second an third term in the representation of equation (\ref{eq6}).}\label{Fig3}
\end{figure}

Next, we derive a representation for $\G_B^{-,+}(\bx,\bxs,\omega)$ in equation (\ref{eq2}).
Substituting the quantities of Table 2 into equation (\ref{eq1}), using equation (\ref{eqex6})
and setting $m=1$ and $n=2$, gives
\begin{equation}\label{eq3}
\G_B^{-,+}(\bx',\bxs,\omega)=\int_{\setdD_1}\R_b^\cup(\bx',\bx,\omega)\G_B^{+,+}(\bx,\bxs,\omega){\rm d}\bx,
\end{equation}
for $\bxs$ just above $\setdD_0$ and $\bx'$ just above $\setdD_1$. Because $\setdD_1$ is transparent (i.e., it does not coincide with an interface), 
equation (\ref{eq3}) does not alter if we take $\bx'$ at $\setdD_1$ instead of just above it. Thus, taking  $\bx'$ at $\setdD_1$, substituting equation (\ref{eq3}) into equation (\ref{eq2})
(with $\bx$ in equation (\ref{eq2})  replaced by $\bx'$), we obtain
\begin{eqnarray}\label{eq4}
&&\hspace{-1.2cm}\R_B^\cup(\bxr,\bxs,\omega)=\R_A^\cup(\bxr,\bxs,\omega)
+\int_{\setdD_1}\int_{\setdD_1}\T_A^-(\bxr,\bx',\omega)\R_b^\cup(\bx',\bx,\omega)\G_B^{+,+}(\bx,\bxs,\omega){\rm d}\bx{\rm d}\bx',
\end{eqnarray}
for $\bxs$ and $\bxr$ just above $\setdD_0$. This is the sought representation for $\R_B^\cup$. In a similar way we find the following representation for $\R_C^\cup$
\begin{eqnarray}\label{eq5}
&&\hspace{-1.2cm}\R_C^\cup(\bxr,\bxs,\omega)=\R_B^\cup(\bxr,\bxs,\omega)
+\int_{\setdD_2}\int_{\setdD_2}\T_B^-(\bxr,\bx',\omega)\R_c^\cup(\bx',\bx,\omega)\G_C^{+,+}(\bx,\bxs,\omega){\rm d}\bx{\rm d}\bx',
\end{eqnarray}
or, upon substitution of equation (\ref{eq4}),
\begin{eqnarray}\label{eq6}
&&\hspace{-.4cm}\R_C^\cup(\bxr,\bxs,\omega)=\R_A^\cup(\bxr,\bxs,\omega)\\
&&\hspace{-.4cm}+\int_{\setdD_1}\int_{\setdD_1}\T_A^-(\bxr,\bx',\omega)\R_b^\cup(\bx',\bx,\omega)\G_B^{+,+}(\bx,\bxs,\omega){\rm d}\bx{\rm d}\bx'\nonumber\\
&&\hspace{-.4cm}+\int_{\setdD_2}\int_{\setdD_2}\T_B^-(\bxr,\bx',\omega)\R_c^\cup(\bx',\bx,\omega)\G_C^{+,+}(\bx,\bxs,\omega){\rm d}\bx{\rm d}\bx',\nonumber
\end{eqnarray}
for $\bxs$ and $\bxr$ just above $\setdD_0$.
The first term on the right-hand side is the reflection response of the overburden (Figure \ref{Fig2}, medium $A$ (= $a$)). The second and third terms on the right-hand side 
contain the reflection responses of the target zone and the underburden\rev{, respectively} (media $b$ and $c$ in Figure \ref{Fig2}). These terms 
are visualised in Figure \ref{Fig3}.

Note that\rev{, if}  the subsurface would be divided into more and thinner units, the recursive derivation process could be continued, 
leading to additional  terms on the right-hand side of equation (\ref{eq6}). In the limiting case (for infinitesimally thin units), 
the reflection responses under the integrals could be replaced by local reflection operators, the Green's functions $\G^{+,+}$ by transmission responses $\T^+$,  
and the sum in the right-hand side would become an integral along the depth coordinate. 
The resulting expression would be the so-called ``generalised  primary representation'' \citep{Kennett74BS, Hubral80GEO, Resnick86GJR, Fishman87JASA, Wapenaar96GJI2, Haines96JMP}.

The representation of equation (\ref{eq6}) is not meant as a recipe for numerical modelling. However, it is a suited starting point for the derivation of a scheme for target replacement.
In equation (\ref{eq6}), $\R_b^\cup(\bx',\bx,\omega)$ represents the reflection response from above of the target zone (unit $b$ in Figure \ref{Fig1}).
Let $\Rbb^\cup(\bx',\bx,\omega)$ denote the reflection response of a changed target zone \rev{(which we denote as unit $\bar b$)}.
The reflection response of the entire medium, with the changed target zone, is given by the following representation
\begin{eqnarray}\label{eq6b}
&&\hspace{-.4cm}\RCb^\cup(\bxr,\bxs,\omega)=\R_A^\cup(\bxr,\bxs,\omega)\\
&&\hspace{-.4cm}+\int_{\setdD_1}\int_{\setdD_1}\T_A^-(\bxr,\bx',\omega)\Rbb^\cup(\bx',\bx,\omega)\GBb^{+,+}(\bx,\bxs,\omega){\rm d}\bx{\rm d}\bx'\nonumber\\
&&\hspace{-.4cm}+\int_{\setdD_2}\int_{\setdD_2}\TBb^-(\bxr,\bx',\omega)\R_c^\cup(\bx',\bx,\omega)\GCb^{+,+}(\bx,\bxs,\omega){\rm d}\bx{\rm d}\bx'.\nonumber
\end{eqnarray}
\rev{Note that, although it is assumed that the overburden and underburden are unchanged, all quantities on the right-hand side  that contain a propagation path through the target zone
 are influenced  by the changes, which is indicated by the bars.}
In the following two sections\rev{,} we discuss the target replacement in detail. 
First, in \rev{S}ection \ref{sec3} we discuss the removal of the target zone response from the original reflection response $\R_C^\cup(\bxr,\bxs,\omega)$. 
Next, in \rev{S}ection \ref{sec4} we discuss how to insert the response of the changed target into the new reflection response $\RCb^\cup(\bxr,\bxs,\omega)$.
 
\section{Removing the target zone from the original reflection response}\label{sec3} 

\begin{figure}
\centerline{\epsfysize=7 cm\epsfbox{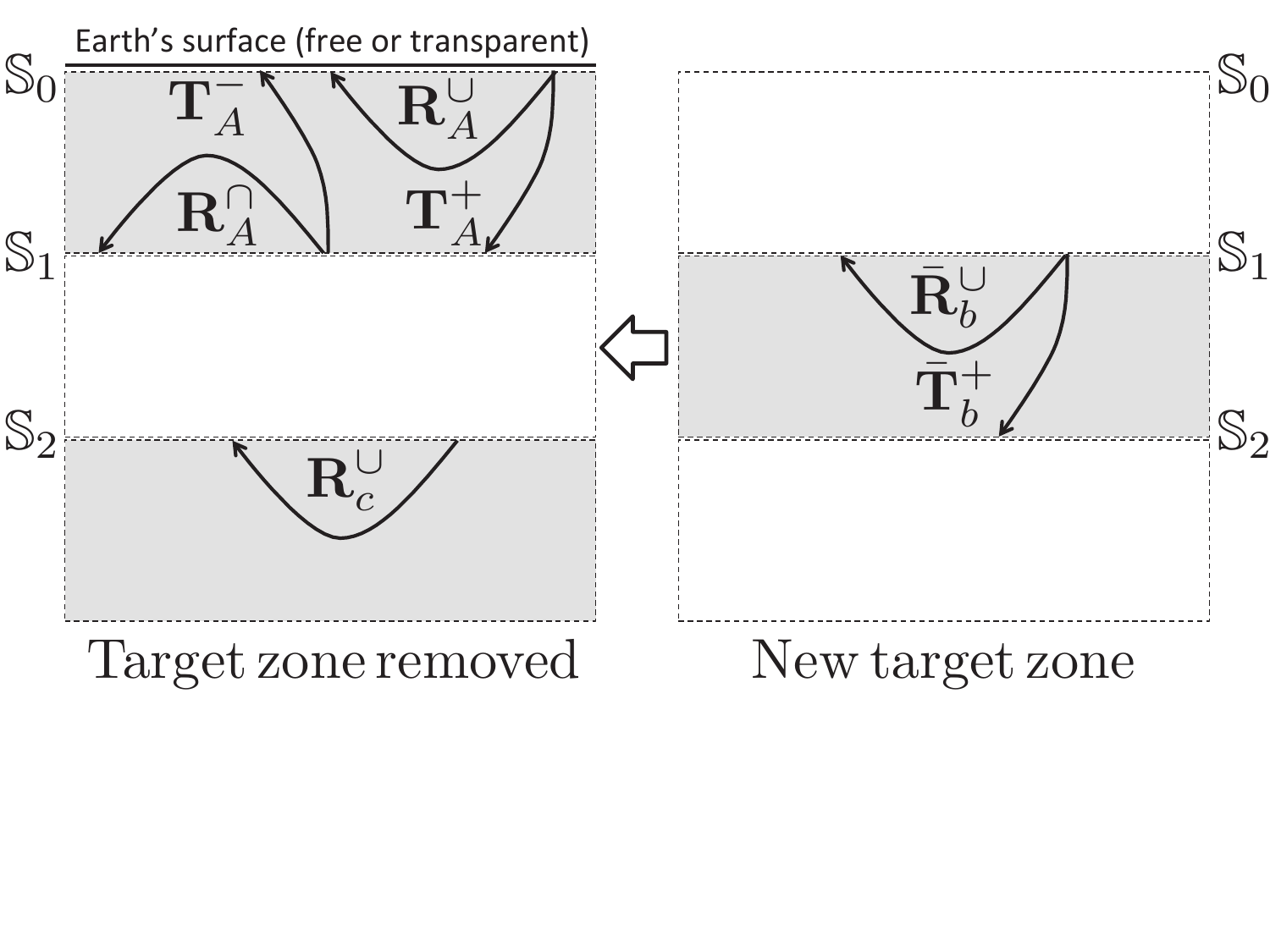}}
\vspace{-1.8cm}
\caption{Left: overburden and underburden responses, obtained from the reflection response $\R_C^\cup$, using the Marchenko method.
Right: modelled responses of the new target zone, to be inserted between the overburden and underburden responses.}\label{Fig5}
\end{figure}

Given the reflection response of the entire medium,  $\R_C^\cup$, our aim is to resolve the responses of the media $A$ (= $a$) and $c$ (i.e., the overburden and underburden, 
Figure \ref{Fig5}). 
\rev{If} $\R_C^\cup$ contained only primary $P$-wave reflections, we could apply simple time-windowing in the time domain to separate the reflection responses of the different units. 
However, because of multiple scattering (possibly including surface-related multiples) and wave conversion, 
the responses of the different units overlap and cannot be straightforwardly separated by time-windowing. 
Here we show that so-called ``focusing functions'', recently introduced for Marchenko imaging \citep{Wapenaar2013PRL, Slob2014GEO}, can be used to obtain the responses of 
media $A$ (= $a$) and $c$.

\begin{figure}
\centerline{\epsfysize=7 cm\epsfbox{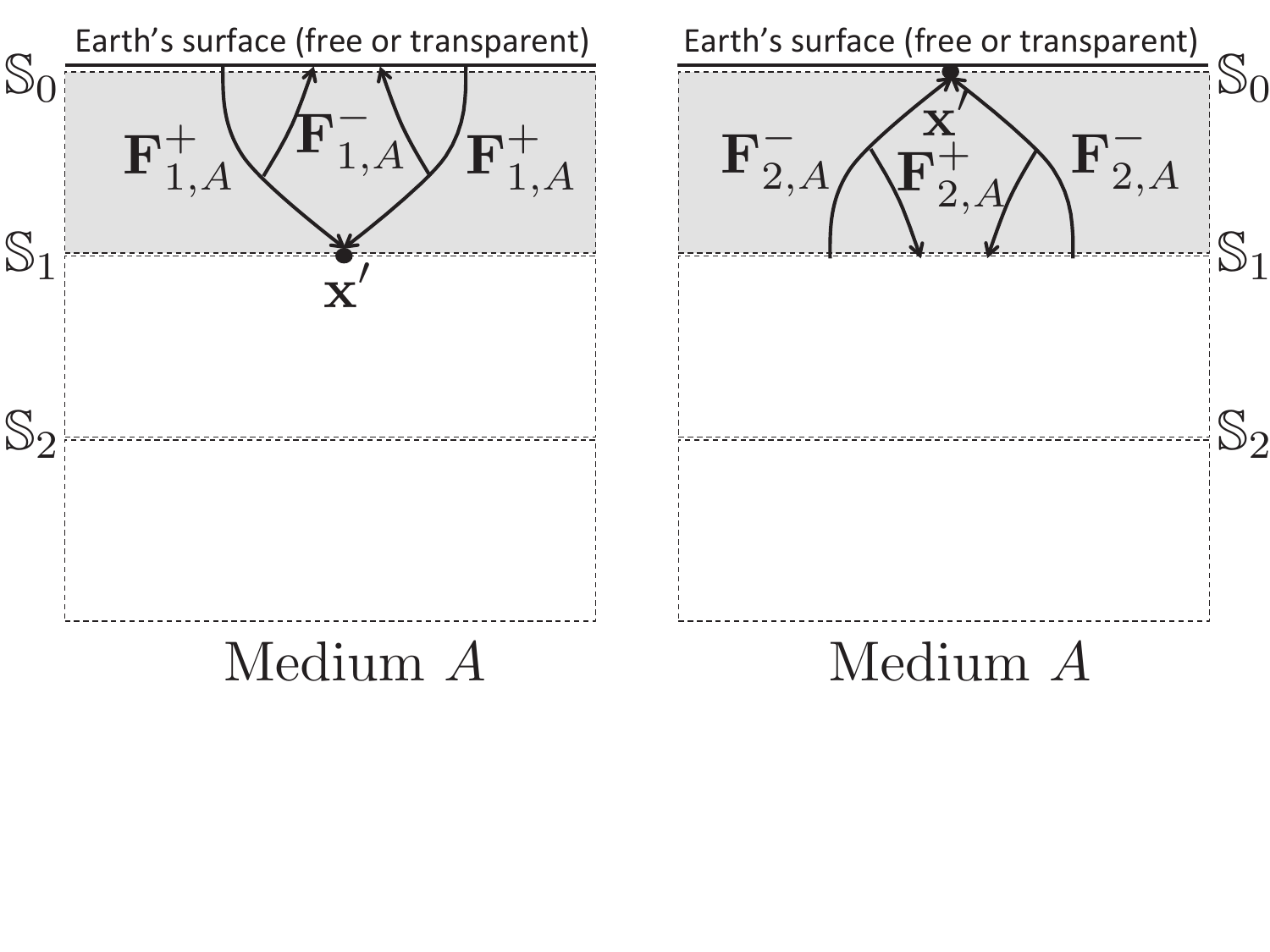}}
\vspace{-1.8cm}
\caption{Focusing functions $\f^\pm(\bx,\bx',\omega)$ and $\ff^\pm(\bx,\bx',\omega)$ in medium $A$.
The rays stand for the full focusing functions, including all orders of multiple scattering and, in the elastodynamic case, mode conversion.}\label{Fig4}
\end{figure}

We start by defining the focusing function $\f^+(\bx,\bx',\omega)$  in medium $A$, with or without free surface just above $\setdD_0$ (Figure \ref{Fig4}). 
Here\rev{,} $\bx'$ defines a focal point at boundary $\setdD_1$, i.e., the lower boundary of unit $a$. 
Hence, $\bx'=(x_1',x_2',x_{3,1})$, with $x_{3,1}$ denoting the depth of $\setdD_1$. The coordinate
$\bx$ is a variable in medium $A$. The superscript $+$ refers to the propagation direction at $\bx$ (which is downgoing in this case).
The focusing function is emitted from all $\bx$ at $\setdD_0$ into medium $A$. Due to scattering in the inhomogeneous medium
 it gives rise to an upgoing function $\f^-(\bx,\bx',\omega)$.
 The focusing conditions for $\bx$ at $\setdD_1$ can be formulated as
\begin{eqnarray}
\hspace{.2cm}\{\f^+(\bx,\bx',\omega)\}_{x_3=x_{3,1}}&=&\I\delta(\bxh-\bxh'),\label{eq25}\\
\{\f^-(\bx,\bx',\omega)\}_{x_3=x_{3,1}}&=&\O,\label{eq26}
\end{eqnarray}
with $\bxh'=(x_1',x_2')$.   Equation (\ref{eq25}) defines the convergence of $\f^+(\bx,\bx',\omega)$ to the focal point  $\bx'$ at $\setdD_1$, whereas equation (\ref{eq26}) states that the
focusing function contains no upward
scattered components at $\setdD_1$, because for medium $A$ the half-space below this boundary is homogeneous. In practical situations evanescent waves are neglected to avoid instability of the focusing
function, hence, the delta function in equation (\ref{eq25}) should be interpreted as a band-limited spatial impulse.

The focusing functions $\f^+(\bx,\bx',\omega)$ and $\f^-(\bx,\bx',\omega)$ for $\bx$ at $\setdD_0$ and $\bx'$ at $\setdD_1$ 
can be obtained from the reflection response $\R_C^\cup(\bxr,\bx,\omega)$ for $\bxr$ just above $\setdD_0$,
using the Marchenko method. We only outline the main features.
In Appendix \ref{AppD}\rev{,} the following relations between  $\R_C^\cup(\bxr,\bx,\omega)$,  $\f^\pm(\bx,\bx',\omega)$  and $\G_C^{\pm,+}(\bx',\bxr,\omega)$ are derived
\begin{eqnarray}\label{eq72kk}
\hspace{.0cm}&&\{\G_C^{-,+}(\bx',\bxr,\omega)\}^t+\f^-(\bxr,\bx',\omega)
=\int_{\setdD_0}\R_C^\cup(\bxr,\bx,\omega)\f^+(\bx,\bx',\omega){\rm d}\bx,
\end{eqnarray}
and
\begin{eqnarray}\label{eq73kka}
\hspace{.0cm}&&\{\G_C^{+,+}(\bx',\bxr,\omega)\}^t-\{\f^+(\bxr,\bx',\omega)\}^*
=-\int_{\setdD_0}\R_C^\cup(\bxr,\bx,\omega)\{\f^-(\bx,\bx',\omega)\}^*{\rm d}\bx,
\end{eqnarray}
(with $\bxr$ just above $\setdD_0$ and $\bx'$ at $\setdD_1$) for the situation that the earth's surface is transparent.
For the acoustic case\rev{,} these equations can be solved for $\f^+(\bx,\bx',\omega)$ and $\f^-(\bx,\bx',\omega)$ using the multidimensional Marchenko method
 \citep{Wapenaar2014GEO, Slob2014GEO, Neut2015GJI,Ravasi2016GJI}. 
The main assumption is that, in addition to $\R_C^\cup(\bxr,\bx,\omega)$, an estimate of the direct arrival of $\f^+(\bx,\bx',\omega)$ 
is available. \rev{This can be defined in a smooth model of the overburden.} The Marchenko method uses causality arguments to separate the Green's functions from the focusing functions in the left-hand sides of 
the time-domain versions of equations (\ref{eq72kk}) and (\ref{eq73kka}).
The multidimensional Marchenko method also holds for the elastodynamic case, 
except that in this case an estimate of the direct arrival plus the forward scattered events of  $\f^+(\bx,\bx',\omega)$  needs to be available \citep{Wapenaar2014GJI}.

For the situation that the earth's surface is a free surface, equations (\ref{eq72kk}) and (\ref{eq73kka}) have been modified by 
\citet{Singh2017GEO}, \rev{\citet{Slob2017EAGE} and \citet{Ravasi2017GEO}}, to account for the surface-related
multiple reflections. In \rev{these approaches,} the surface-related multiples are present in the reflection response, but not in the focusing functions. 
For the target replacement procedure discussed in this paper it is more convenient to use focusing functions that include surface-related multiples. 
From the derivation in Appendix \ref{AppD} it follows that for this situation equation (\ref{eq72kk}) remains valid (but 
with all quantities now including the surface-related multiples), and that
equation (\ref{eq73kka}) needs to be replaced by
\begin{eqnarray}\label{eq73kkb}
&&\hspace{-.7cm}\{\G_C^{+,+}(\bx',\bxr,\omega)\}^t-\{\f^+(\bxr,\bx',\omega)+\r\f^-(\bxr,\bx',\omega)\}^*\nonumber\\
&&\hspace{-.7cm}=\int_{\setdD_0}\R_C^\cup(\bxr,\bx,\omega)\r\{\f^+(\bx,\bx',\omega)\}^*{\rm d}\bx
\end{eqnarray}
(with $\bxr$ just above $\setdD_0$ and $\bx'$ at $\setdD_1$). The set of equations (\ref{eq72kk}) and (\ref{eq73kkb}) for the situation with free surface can be solved in a similar way as the 
set of equations (\ref{eq72kk}) and (\ref{eq73kka}) for the situation without free surface. A further discussion of the multidimensional Marchenko method
to resolve $\f^\pm(\bx,\bx',\omega)$ from the reflection response $\R_C^\cup(\bxr,\bx,\omega)$ is beyond the scope of this paper.

Assuming the focusing functions $\f^+(\bx,\bx',\omega)$ and $\f^-(\bx,\bx',\omega)$ have been found, we use these to resolve the responses of medium $A$.
In Appendix \ref{AppA}\rev{,} we show that the  response to focusing function $\f^+(\bx,\bx',\omega)$, when emitted from $\setdD_0$ into medium $A$, can be quantified as follows 
\begin{equation}\label{eq70kk}
\I\delta(\bxh''-\bxh')=\int_{\setdD_0}\T_A^+(\bx'',\bx,\omega)\f^+(\bx,\bx',\omega){\rm d}\bx,
\end{equation}
for $\bx'$  and  $\bx''$ at $\setdD_1$, and
\begin{equation}\label{eq33kk}
\f^-(\bxr,\bx',\omega)=\int_{\setdD_0}\R_A^\cup(\bxr,\bx,\omega)\f^+(\bx,\bx',\omega){\rm d}\bx,
\end{equation}
for $\bxr$ just above $\setdD_0$ and $\bx'$ at $\setdD_1$.
Equation (\ref{eq70kk}) describes the transmission response of medium $A$ to the focusing function.
The response at $\setdD_1$ is a (band-limited) spatial impulse (consistent with the focusing condition of equation (\ref{eq25})). 
Equation (\ref{eq33kk}) describes the reflection response of medium $A$ to the focusing function.
The response at $\setdD_0$ is the upgoing part of the focusing function. 
Both equations (\ref{eq70kk}) and (\ref{eq33kk}) hold for the situation with or without free surface just above $\setdD_0$.
Inverting these equations yields the transmission response $\T_A^+(\bx'',\bx,\omega)$ (which, according to equation (\ref{eq70kk}) is the inverse of the focusing function $\f^+(\bx,\bx',\omega)$)
and the reflection response $\R_A^\cup(\bxr,\bx,\omega)$ of medium $A$, the overburden (Figure \ref{Fig5}). 

To derive the response of medium $A$ from below, we introduce a second focusing function
 $\ff^-(\bx,\bx',\omega)$  in medium $A$, with or without free surface just above $\setdD_0$ (Figure \ref{Fig4}). 
This time $\bx'$ defines a focal point at boundary $\setdD_0$, i.e., the upper boundary of unit $a$. 
Hence, $\bx'=(x_1',x_2',x_{3,0})$, with $x_{3,0}$ denoting the depth of $\setdD_0$. The coordinate
$\bx$ is a variable in medium $A$. The superscript $-$ refers to the propagation direction at $\bx$ (which is upgoing in this case).
The focusing function is emitted from all $\bx$ at $\setdD_1$ into medium $A$. 
Due to scattering in the inhomogeneous medium\rev{,} 
it gives rise to a downgoing function $\ff^+(\bx,\bx',\omega)$.
 The focusing conditions for $\bx$ at $\setdD_0$ can be formulated as
\begin{eqnarray}
\hspace{0.cm}\{\ff^-(\bx,\bx',\omega)\}_{x_3=x_{3,0}}&=&\I\delta(\bxh-\bxh'),\label{eq25a}\\
\{\ff^+(\bx,\bx',\omega)\}_{x_3=x_{3,0}}&=&\r\I\delta(\bxh-\bxh').\label{eq26a}
\end{eqnarray}
Equation (\ref{eq25a}) defines the convergence of $\ff^-(\bx,\bx',\omega)$ to the focal point  $\bx'$ at $\setdD_0$, 
whereas equation (\ref{eq26a}) accounts for the downward reflection of the upgoing focusing function at  $\setdD_0$. This term vanishes when the earth's surface is transparent.
In Appendix \ref{AppB}\rev{,} we show that the  response to focusing function $\ff^-(\bx,\bx',\omega)$, when emitted from $\setdD_1$ into medium $A$, can be quantified as follows 
\begin{equation}\label{eq70ll}
\I\delta(\bxh''-\bxh')=\int_{\setdD_1}\T_A^-(\bx'',\bx,\omega)\ff^-(\bx,\bx',\omega){\rm d}\bx,
\end{equation}
for $\bx'$  and  $\bx''$ at $\setdD_0$, and
\begin{equation}\label{eq333ll}
\ff^+(\bx'',\bx',\omega)=\int_{\setdD_1}\R_A^\cap(\bx'',\bx,\omega)\ff^-(\bx,\bx',\omega){\rm d}\bx,
\end{equation}
for  $\bx''$ just below $\setdD_1$ and $\bx'$ at $\setdD_0$.
Inverting these equations yields the transmission response $\T_A^-(\bx'',\bx,\omega)$ (which, according to equation (\ref{eq70ll}) is the inverse of the focusing function $\ff^-(\bx,\bx',\omega)$)
and the reflection response $\R_A^\cap(\bx'',\bx,\omega)$ of medium $A$ from below (Figure \ref{Fig5}). 
In Appendix \ref{AppC} we show that the focusing functions $\ff^+$ and $\ff^-$  are related to the focusing functions $\f^+$ and $\f^-$, according to
\begin{equation}\label{eq55kk}
\f^+(\bx'',\bx',\omega)=\{\ff^-(\bx',\bx'',\omega)\}^t,
\end{equation}
and 
\begin{equation}\label{eq56akk}
\f^-(\bx'',\bx',\omega)=-\{\ff^+(\bx',\bx'',\omega)\}^\dagger
\end{equation} 
(with $\bx''$ at $\setdD_0$ and $\bx'$ at $\setdD_1$) for the situation that the earth's surface is transparent.
For the situation that the earth's surface is a free surface, equation (\ref{eq55kk}) remains valid, and equation 
(\ref{eq56akk}) needs to be replaced by
\begin{equation}\label{eq56bkk}
(\r)^*\f^+(\bx'',\bx',\omega)=\{\ff^+(\bx',\bx'',\omega)\}^\dagger
\end{equation} 
(with $\bx''$ at $\setdD_0$ and $\bx'$ at $\setdD_1$).

Next we discuss how to obtain the response of unit $c$, the underburden, from $\R_C^\cup$.
We consider again equations (\ref{eq72kk}) and (\ref{eq73kka}) (or (\ref{eq73kkb})),
this time with $\bx'$ at $\setdD_2$ and $\f^\pm(\bx,\bx',\omega)$ replaced by $\fb^\pm(\bx,\bx',\omega)$.
The focusing functions in medium $B$ can be obtained from the reflection response $\R_C^\cup(\bxr,\bx,\omega)$, 
using the multidimensional Marchenko method, under the same assumptions as outlined above.
Once these focusing functions have been found, they can be substituted into \rev{the modified} equations (\ref{eq72kk}) and (\ref{eq73kka}) (or (\ref{eq73kkb})), yielding the Green's
functions $\G_C^{\pm,+}(\bx',\bxr,\omega)$, with $\bxr$ just above $\setdD_0$ and $\bx'$ at $\setdD_2$.
Analogous to equation (\ref{eq3}), 
these Green's function are mutually related via
\begin{equation}\label{eq3mm}
\G_C^{-,+}(\bx',\bxr,\omega)=\int_{\setdD_2}\R_c^\cup(\bx',\bx,\omega)\G_C^{+,+}(\bx,\bxr,\omega){\rm d}\bx.
\end{equation}
Inversion of equation (\ref{eq3mm}) yields the reflection response $\R_c^\cup(\bx',\bx,\omega)$ for $\bx$ and $\bx'$ at $\setdD_2$ (Figure \ref{Fig5}).

We summarise the steps discussed in this section. Starting with the  reflection response of the entire medium, $\R_C^\cup(\bxr,\bx,\omega)$, 
use the Marchenko method to derive the focusing functions
$\f^\pm(\bx,\bx',\omega)$ and $\ff^\pm(\bx,\bx',\omega)$ for medium $A$. Resolve the responses of the overburden, $\T_A^+(\bx'',\bx,\omega)$,
$\R_A^\cup(\bxr,\bx,\omega)$, $\T_A^-(\bx'',\bx,\omega)$ and $\R_A^\cap(\bx'',\bx,\omega)$,   by inverting equations (\ref{eq70kk}), (\ref{eq33kk}), 
(\ref{eq70ll}) and (\ref{eq333ll}). Next, use the Marchenko method to derive 
the Green's functions $\G_C^{\pm,+}(\bx',\bxr,\omega)$, for $\bx'$ at $\setdD_2$. Resolve the reflection response of the underburden, $\R_c^\cup(\bx',\bx,\omega)$,
by inverting equation (\ref{eq3mm}). The resolved responses are free of an imprint of unit $b$,  the target zone.

\section{Inserting a new target zone into the reflection response}\label{sec4} 

Given the retrieved responses of the overburden 
(medium $A$) 
and underburden 
(unit $c$) 
and a model of the changed target zone
 (unit $\bar b$), 
our aim is to obtain the reflection response $\RCb^\cup(\bxr,\bxs,\omega)$ of the entire medium with the new target zone
 (medium $\bar C$). 
The procedure starts by numerically modelling the reflection and transmission responses of the new target zone, $\Rbb^\cup(\bx',\bx,\omega)$ and 
$\Tbb^+(\bx',\bx,\omega)$ (Figure \ref{Fig5}).
Next, the response $\RCb^\cup(\bxr,\bxs,\omega)$ is built up step by step, using  equation (\ref{eq6b}) as the underlying representation.
Analogous to equations (\ref{eq4}) and (\ref{eq5}), we rewrite equation (\ref{eq6b}) as a cascade of two representations, as follows
\begin{eqnarray}\label{eq4ag}
&&\hspace{-1.2cm}\RBb^\cup(\bxr,\bxs,\omega)=\R_A^\cup(\bxr,\bxs,\omega)
+\int_{\setdD_1}\int_{\setdD_1}\T_A^-(\bxr,\bx',\omega)\Rbb^\cup(\bx',\bx,\omega)\GBb^{+,+}(\bx,\bxs,\omega){\rm d}\bx{\rm d}\bx',
\end{eqnarray}
followed by
\begin{eqnarray}\label{eq5ag}
&&\hspace{-1.2cm}\RCb^\cup(\bxr,\bxs,\omega)=\RBb^\cup(\bxr,\bxs,\omega)
+\int_{\setdD_2}\int_{\setdD_2}\TBb^-(\bxr,\bx',\omega)\R_c^\cup(\bx',\bx,\omega)\GCb^{+,+}(\bx,\bxs,\omega){\rm d}\bx{\rm d}\bx',
\end{eqnarray}
for $\bxs$ and $\bxr$ just above $\setdD_0$.
Quantities in these representations that still need to be determined are $\GBb^{+,+}(\bx,\bxs,\omega)$, 
$\GCb^{+,+}(\bx,\bxs,\omega)$ and $\TBb^-(\bxr,\bx',\omega)$.

In Appendix \ref{AppE}\rev{,} we derive the following equation for the unknown $\GBb^{+,+}(\bx,\bxs,\omega)$
\begin{equation}\label{eq99bb}
\T_A^+(\bx'',\bxs,\omega)=\int_{\setdD_1}\MAbb(\bx'',\bx,\omega)\GBb^{+,+}(\bx,\bxs,\omega){\rm d}\bx,
\end{equation}
with
\begin{eqnarray}\label{eq999bb}
\hspace{.5cm}\MAbb(\bx'',\bx,\omega)&=&\I\delta(\bxh''-\bxh)
-\int_{\setdD_1}\R_A^\cap(\bx'',\bx',\omega)\Rbb^\cup(\bx',\bx,\omega){\rm d}\bx', 
\end{eqnarray}
for $\bxs$ just above $\setdD_0$, and $\bx$ and  $\bx''$ at $\setdD_1$.
Since $\T_A^+$, $\R_A^\cap$ and $\Rbb^\cup$ are known, $\GBb^{+,+}(\bx,\bxs,\omega)$ can be resolved
by inverting equation (\ref{eq99bb}). Substituting  this into equation (\ref{eq4ag}), together with the other quantities that are already known, yields $\RBb^\cup(\bxr,\bxs,\omega)$.

Similarly $\GCb^{+,+}(\bx,\bxs,\omega)$ can be resolved by inverting 
\begin{equation}\label{eq99bbcc}
\TBb^+(\bx'',\bxs,\omega)=\int_{\setdD_2}\MBcb(\bx'',\bx,\omega)\GCb^{+,+}(\bx,\bxs,\omega){\rm d}\bx,
\end{equation}
with
\begin{eqnarray}\label{eq999bbcc}
\hspace{.5cm}\MBcb(\bx'',\bx,\omega)&=&\I\delta(\bxh''-\bxh)
-\int_{\setdD_2}\RBb^\cap(\bx'',\bx',\omega)\R_c^\cup(\bx',\bx,\omega){\rm d}\bx', 
\end{eqnarray}
for $\bxs$ just above $\setdD_0$, and $\bx$ and  $\bx''$ at $\setdD_2$. This requires expressions for $\TBb^+(\bx'',\bxs,\omega)$ and $\RBb^\cap(\bx'',\bx',\omega)$.

In Appendix \ref{AppF} we derive the following representation for $\TBb^+(\bx'',\bxs,\omega)$
\begin{equation}\label{eq15mm}
\TBb^+(\bx'',\bxs,\omega)=\int_{\setdD_1}\Tbb^+(\bx'',\bx,\omega)\GBb^{+,+}(\bx,\bxs,\omega){\rm d}\bx,
\end{equation}
for $\bxs$ just above $\setdD_0$ and  $\bx''$ at $\setdD_2$. Note that $\TBb^-(\bxr,\bx',\omega)$, needed in equation (\ref{eq5ag}), follows by applying equation (\ref{eqex14}).

In Appendix \ref{AppG}\rev{,} we derive the following equation for the unknown $\RBb^\cap(\bx,\bx',\omega)$
\begin{eqnarray}
\hspace{.0cm}&&\int_{\setdD_2}\{\TBb^-(\bxs,\bx,\omega)\}^*\RBb^\cap(\bx,\bx',\omega){\rm d}\bx=
-\int_{\setdD_0}\{\RBb^\cup(\bxs,\bx,\omega)\}^*\TBb^-(\bx,\bx',\omega){\rm d}\bx,\label{eq101bc}
\end{eqnarray}
(with $\bxs$ just above $\setdD_0$ and  $\bx'$ at $\setdD_2$) for the situation that the earth's surface is transparent.
For the situation that the earth's surface is a free surface, this equation needs to be replaced by
\begin{equation}\label{eq102b}
\int_{\setdD_2}\{\TBb^-(\bxs,\bx,\omega)\}^*\RBb^\cap(\bx,\bx',\omega){\rm d}\bx=\r\TBb^-(\bxs,\bx',\omega),
\end{equation}
(with $\bxs$ just above $\setdD_0$ and  $\bx'$ at $\setdD_2$).
Since  $\RBb^\cup$ and $\TBb^-$ are known,
$\RBb^\cap(\bx,\bx',\omega)$ can be resolved by inverting either equation (\ref{eq101bc}) or (\ref{eq102b}). 

We summarise the steps discussed in this section. Starting with a model of the new target zone, determine its responses $\Rbb^\cup(\bx',\bx,\omega)$ and 
$\Tbb^+(\bx',\bx,\omega)$ by numerical modelling. Next, resolve the Green's function of medium $\bar B$, $\GBb^{+,+}(\bx,\bxs,\omega)$, by inverting equation (\ref{eq99bb}).
Substitute this, together with $\Rbb^\cup(\bx',\bx,\omega)$, into equation (\ref{eq4ag}), which yields the reflection response of medium $\bar B$, $\RBb^\cup(\bxr,\bxs,\omega)$.
Resolve $\RBb^\cap(\bx,\bx',\omega)$ by inverting equation (\ref{eq101bc}) or (\ref{eq102b}). 
Substitute  this into equation (\ref{eq999bbcc})
and, subsequently, substitute the result $\MBcb(\bx'',\bx,\omega)$  into equation (\ref{eq99bbcc}).
Resolve $\GCb^{+,+}(\bx,\bxs,\omega)$ by inverting equation (\ref{eq99bbcc}).
Substitute this, together with the other quantities that are already known, into equation (\ref{eq5ag}), which yields the sought reflection response $\RCb^\cup(\bxr,\bxs,\omega)$.

\begin{figure}
\centerline{\epsfxsize=9 cm\epsfbox{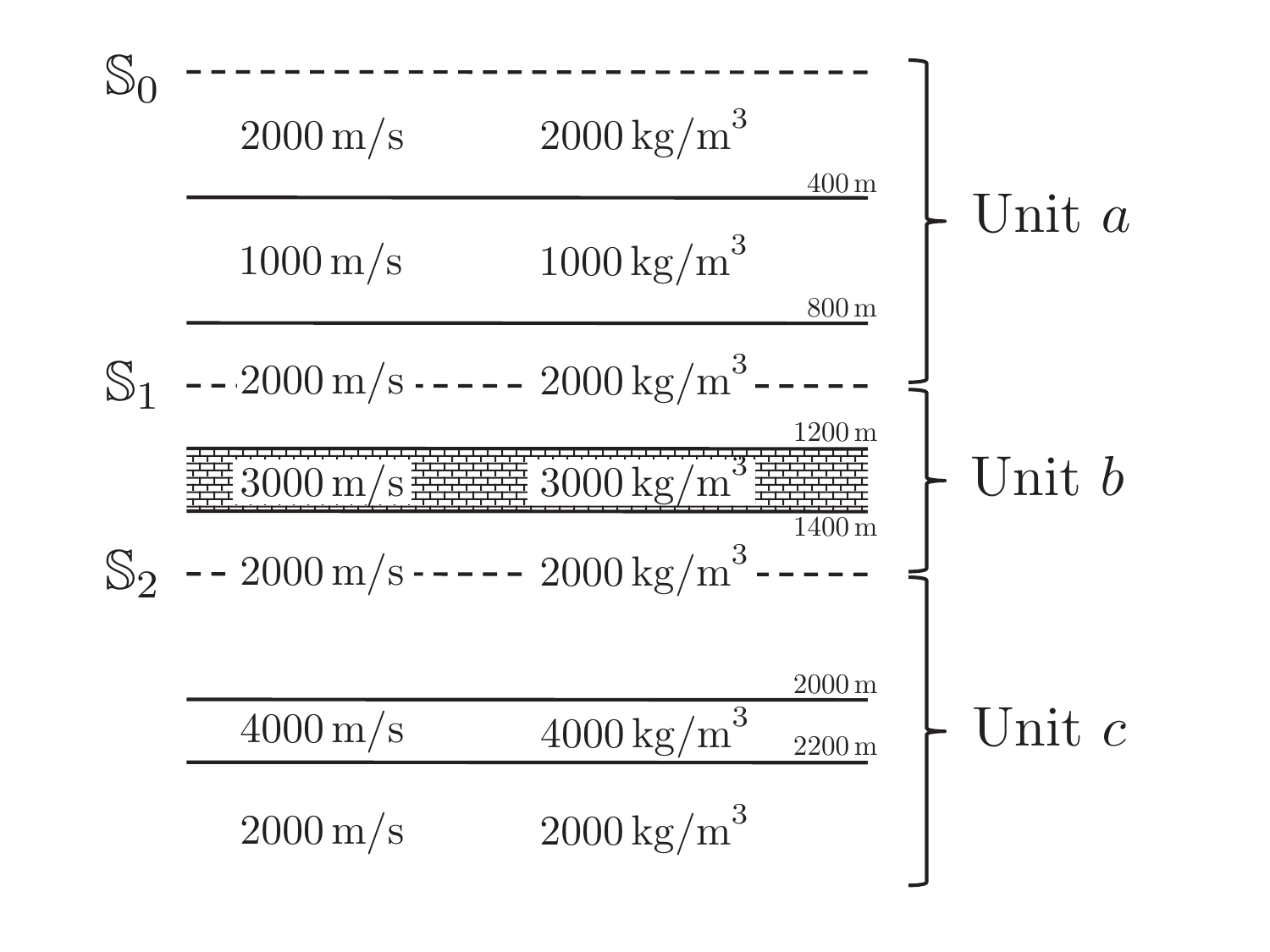}}
\vspace{-.3cm}
\caption{
Horizontally layered medium \rev{for the  plane-wave experiment}, with the three units indicated. The earth's surface is considered transparent.}\label{Fig6}
\end{figure}

\begin{figure}
\centerline{\epsfxsize=9. cm\epsfbox{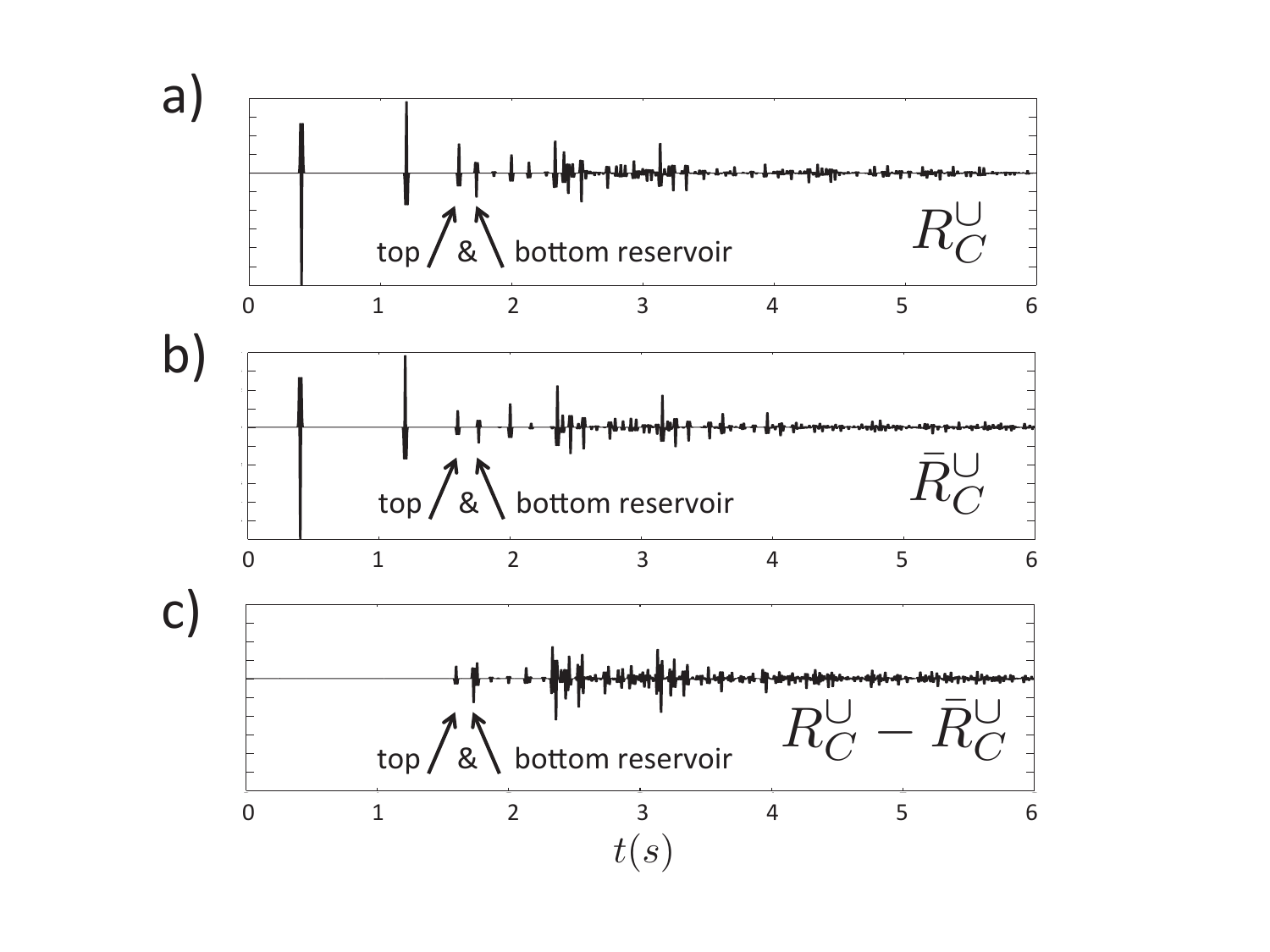}}
\vspace{-.7cm}
\caption{ 
(a) Numerically modelled reflection response of the model of Figure \ref{Fig6}. (b) Numerically modelled time-lapse response.
(c) The difference of the responses in (a) and (b).}\label{Fig7}
\end{figure}

\begin{figure}
\centerline{\epsfxsize=9. cm\epsfbox{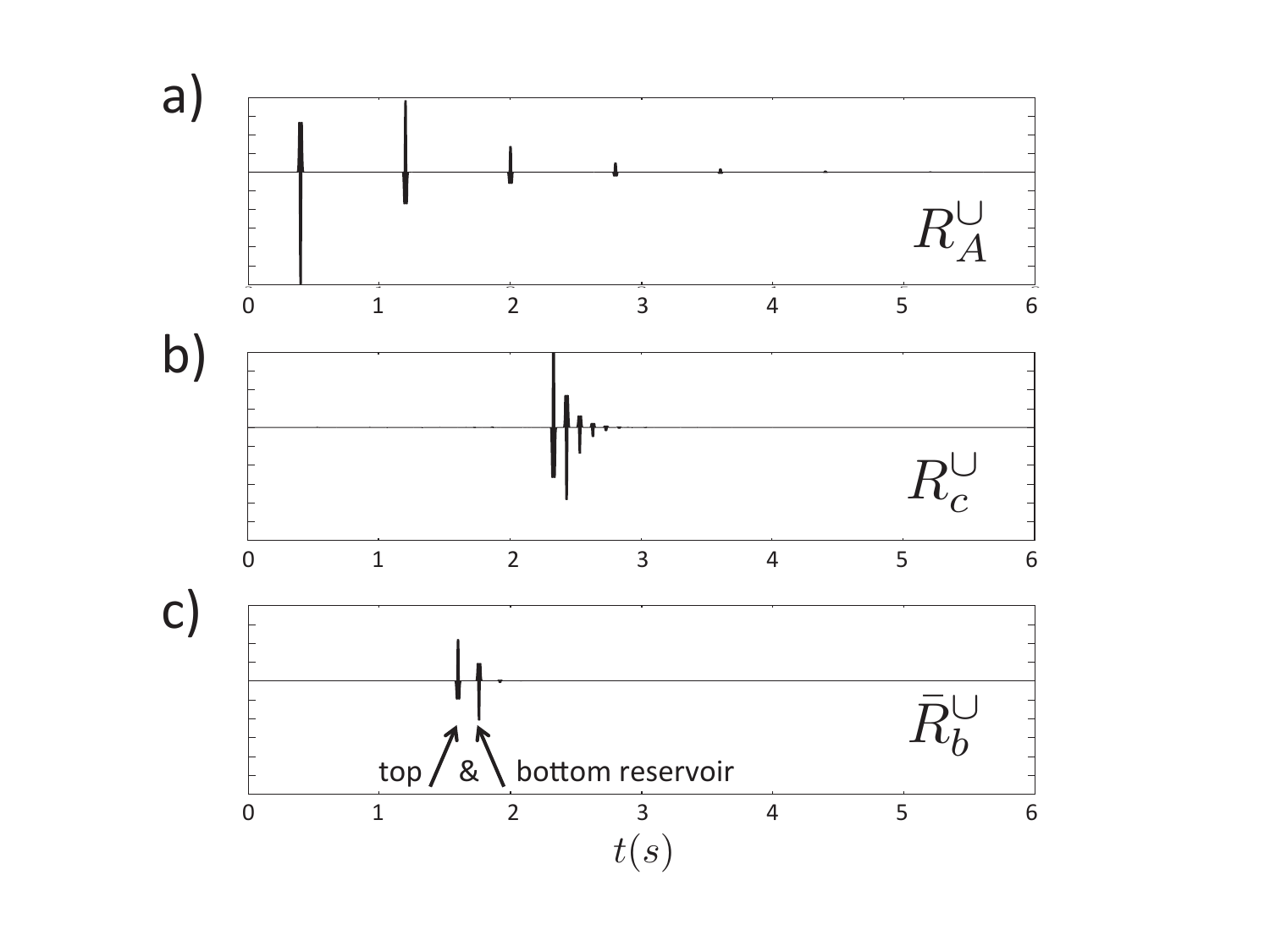}}
\vspace{-.7cm}
\caption{  
(a) The response of medium $A$ (the overburden), retrieved from $R_C^\cup(t)$. (b) The response of unit $c$ (the underburden), retrieved from $R_C^\cup(t)$.
(c) Numerically modelled response of unit $\bar b$ (the new target zone).}\label{Fig8}
\end{figure}

\begin{figure}
\centerline{\epsfxsize=9. cm\epsfbox{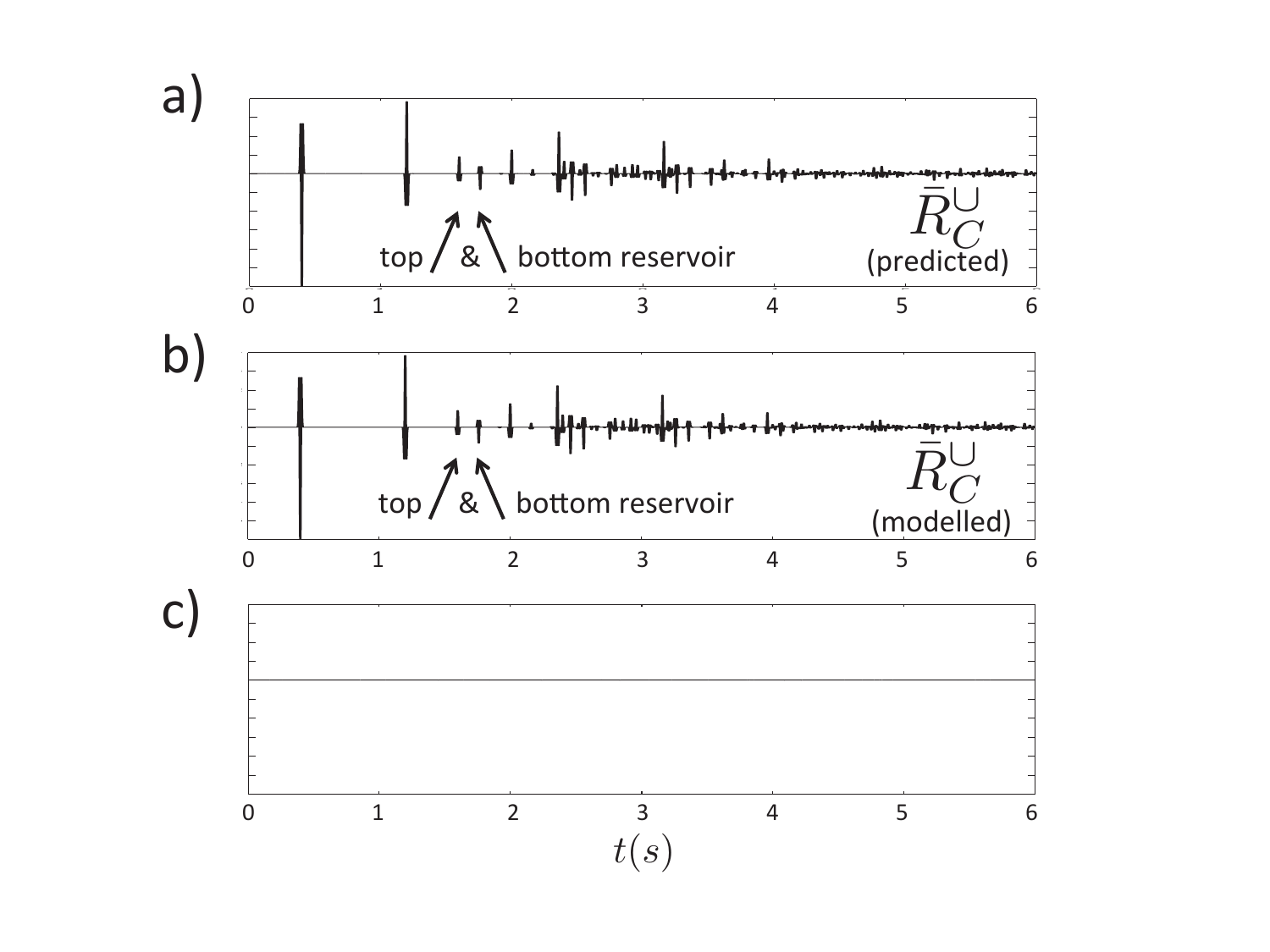}}
\vspace{-.7cm}
\caption{ (a) The predicted time-lapse response $\RC^\cup(t)$, constructed from the responses in Figure \ref{Fig8}.
(b) For comparison, the numerically modelled time-lapse response.
(c) The difference of the responses in (a) and (b).
}\label{Fig9}
\end{figure}

\section{Numerical examples}\label{sec5}

We illustrate the proposed method with two numerical examples. Although the method holds for vertically and laterally inhomogeneous media, for simplicity we consider laterally invariant media 
in the following examples.

 In the first example\rev{,} we consider the acoustic plane-wave response of a horizontally layered medium, 
without free surface (which is the  situation after surface-related multiple elimination). Figure \ref{Fig6} shows the horizontally layered medium. 
The velocities are given in m/s, the mass densities in kg/m$^3$, and the depth of the interfaces (denoted by the solid lines) in m. 
To emphasise internal multiples, the mass densities have the same numerical values as the propagation velocities.
The layer between 1200 m and 1400 m represents a reservoir (hence, this is the layer in which changes will take place). 
The target zone (unit $b$) includes this reservoir layer (the remainder of the target zone will, however, not undergo any changes).
Figure \ref{Fig7}(a) shows the numerically modelled plane-wave reflection response $R_C^\cup(t)$ at $\setdD_0$ in the time domain, convolved with a Ricker wavelet with a central frequency of 50 Hz
(note that we replaced the boldface symbol $\R$ by a plain $R$, because the acoustic response is a scalar function; moreover, we replaced $\omega$ by $t$ because the response is shown in the time domain). 
The reflections from the top and bottom of the reservoir are indicated by arrows. 
We consider a time-lapse scenario, in which the velocity in the reservoir is changed from 3000 m/s to 2500 m/s (and a similar change is applied  to the mass density).
Figure \ref{Fig7}(b) shows the numerically modelled time-lapse reflection response $\RC^\cup(t)$ and Figure \ref{Fig7}(c) shows the difference $R_{C}^\cup(t)-\RC^\cup(t)$.
Note the significant multiple coda, following the difference response of the reservoir. Our aim is to show that the time-lapse response (Figure \ref{Fig7}(b))
can be predicted from the original response (Figure \ref{Fig7}(a)) by target replacement.

Following the procedure discussed in \rev{S}ection \ref{sec3} \rev{(simplified for the 1D situation)}, we remove the response of the target zone from the reflection response $R_C^\cup(t)$.
The overburden response $R_A^\cup(t)$, resolved from equation (\ref{eq33kk}), is shown in the time domain in Figure \ref{Fig8}(a). 
Note that it contains the first two events of  $R_C^\cup(t)$ and a coda due to the internal multiples in the low-velocity layer in the overburden.
The underburden response $R_c^\cup(t)$, resolved from equation (\ref{eq3mm}), is shown in Figure \ref{Fig8}(b). 
For display purposes it has been shifted in time, so that the travel times correspond with those in Figure \ref{Fig7}(a).

Following the procedure discussed in \rev{S}ection \ref{sec4} \rev{(simplified for the 1D situation)}, we predict the time-lapse response. To this end,
we first model the response of the new target zone, $\Rb^\cup(t)$. This is shown in Figure \ref{Fig8}(c).
For display purposes\rev{,} it has been shifted in time so that the travel time to the top of the reservoir corresponds with that in Figure \ref{Fig7}(a).
 The predicted time-lapse reflection response at the surface, $\RC^\cup(t)$, obtained with the representations of equations (\ref{eq4ag}) and (\ref{eq5ag}),
is shown in the time domain in Figure \ref{Fig9}(a). The numerically modelled response of Figure \ref{Fig7}(b), is once more shown (as a reference) in Figure \ref{Fig9}(b).
The difference of the predicted and modelled responses is shown in Figure \ref{Fig9}(c) and appears to be practically zero.
This confirms that the new reflection response $\RC^\cup(t)$ has been very accurately predicted by the proposed method.

\begin{figure}
\centerline{\epsfxsize=9 cm\epsfbox{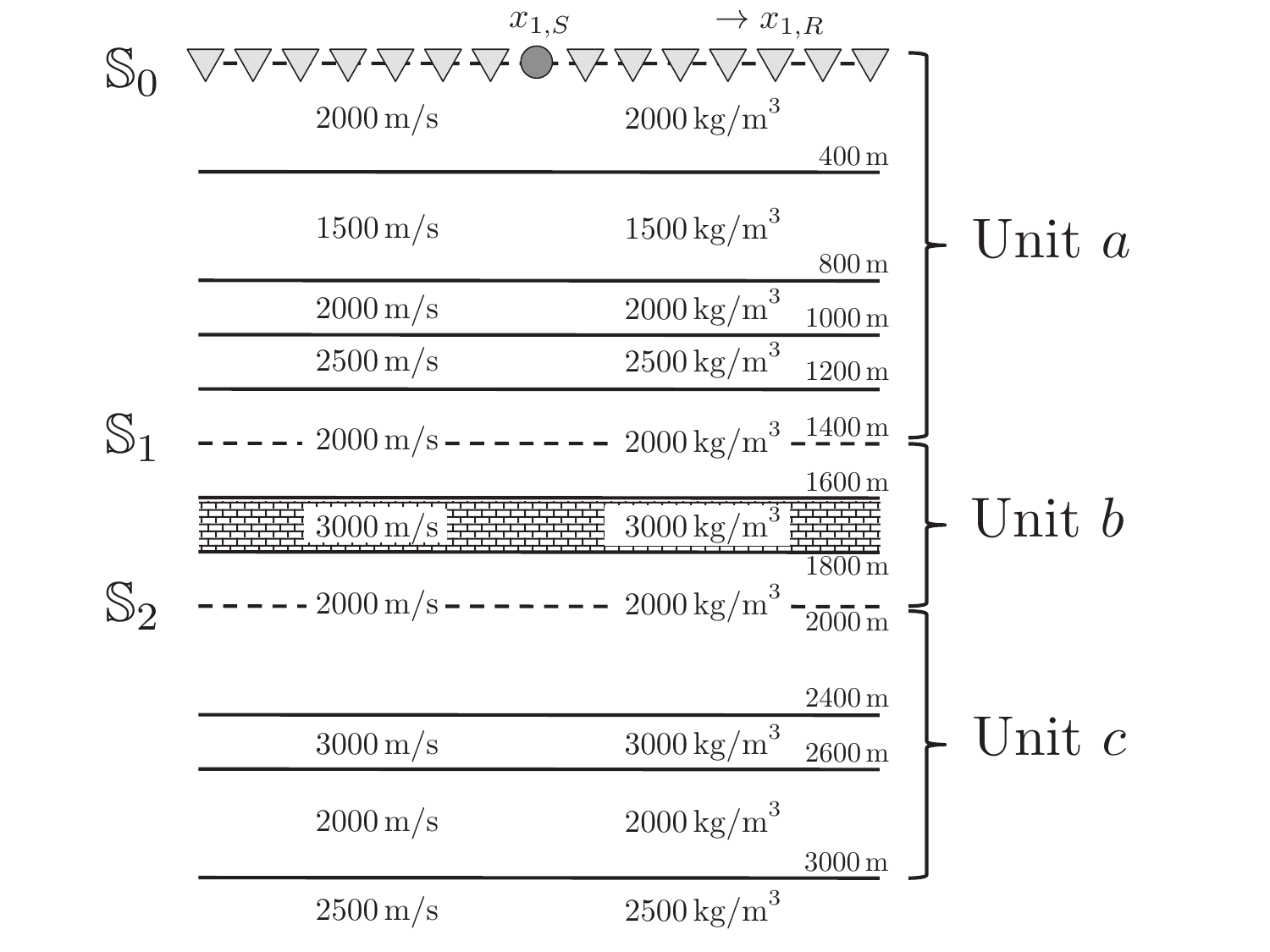}}
\vspace{-.3cm}
\caption{
\rev{Horizontally layered medium for the 2D experiment. }}\label{Fig11a}
\end{figure}

\begin{figure}
\centerline{\epsfxsize=9. cm\epsfbox{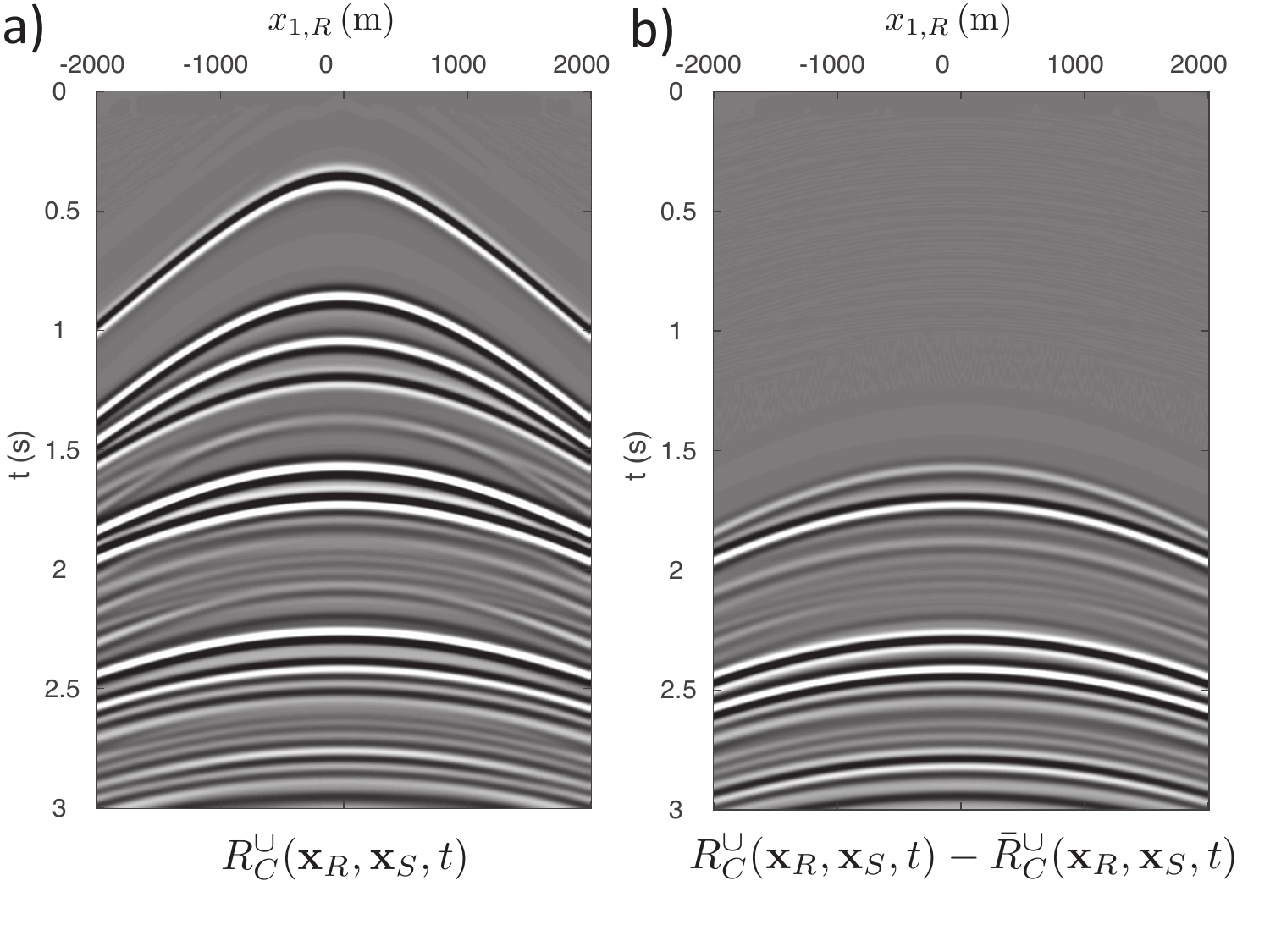}}
\vspace{-.5cm}
\caption{ 
(a) Numerically modelled 2D reflection response. (b) Numerically modelled difference response.}\label{Fig10}
\end{figure}

\begin{figure}
\centerline{\epsfxsize=9. cm\epsfbox{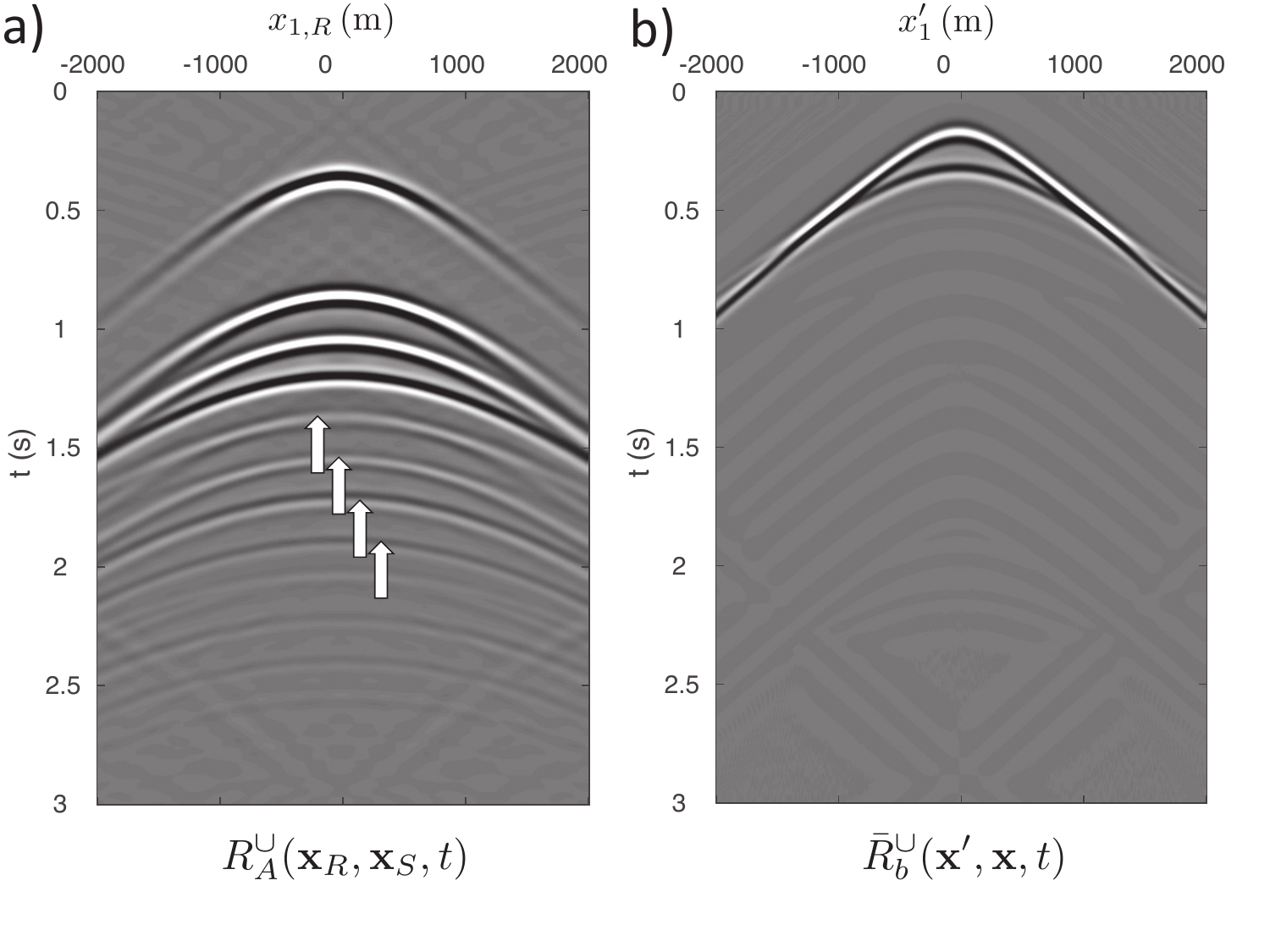}}
\vspace{-.5cm}
\caption{ 
(a) The response of medium $A$ (the overburden), retrieved from $R_C^\cup(\bx_R,\bx_S,t)$. 
(b) Numerically modelled response of the new target zone.}\label{Fig11}
\end{figure}

\begin{figure}
\centerline{\epsfxsize=9. cm\epsfbox{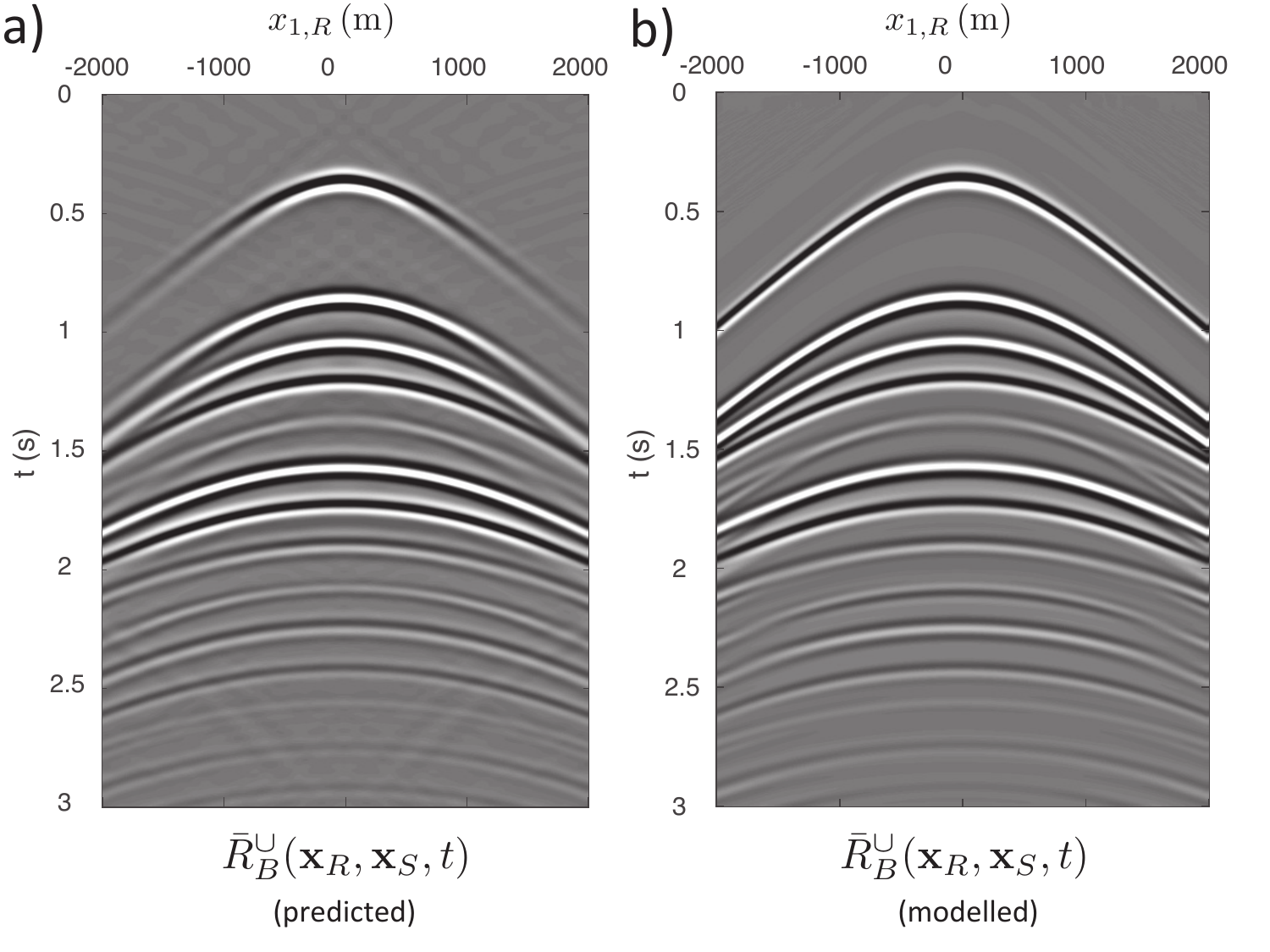}}
\vspace{-.2cm}
\caption{ 
\rev{(a) The predicted time-lapse response $\RB^\cup(\bx_R,\bx_S,t)$, constructed from the responses in Figure \ref{Fig11}.
(b) For comparison, the numerically modelled time-lapse response.}}\label{Fig12a}
\end{figure}

\begin{figure}
\centerline{\epsfxsize=9. cm\epsfbox{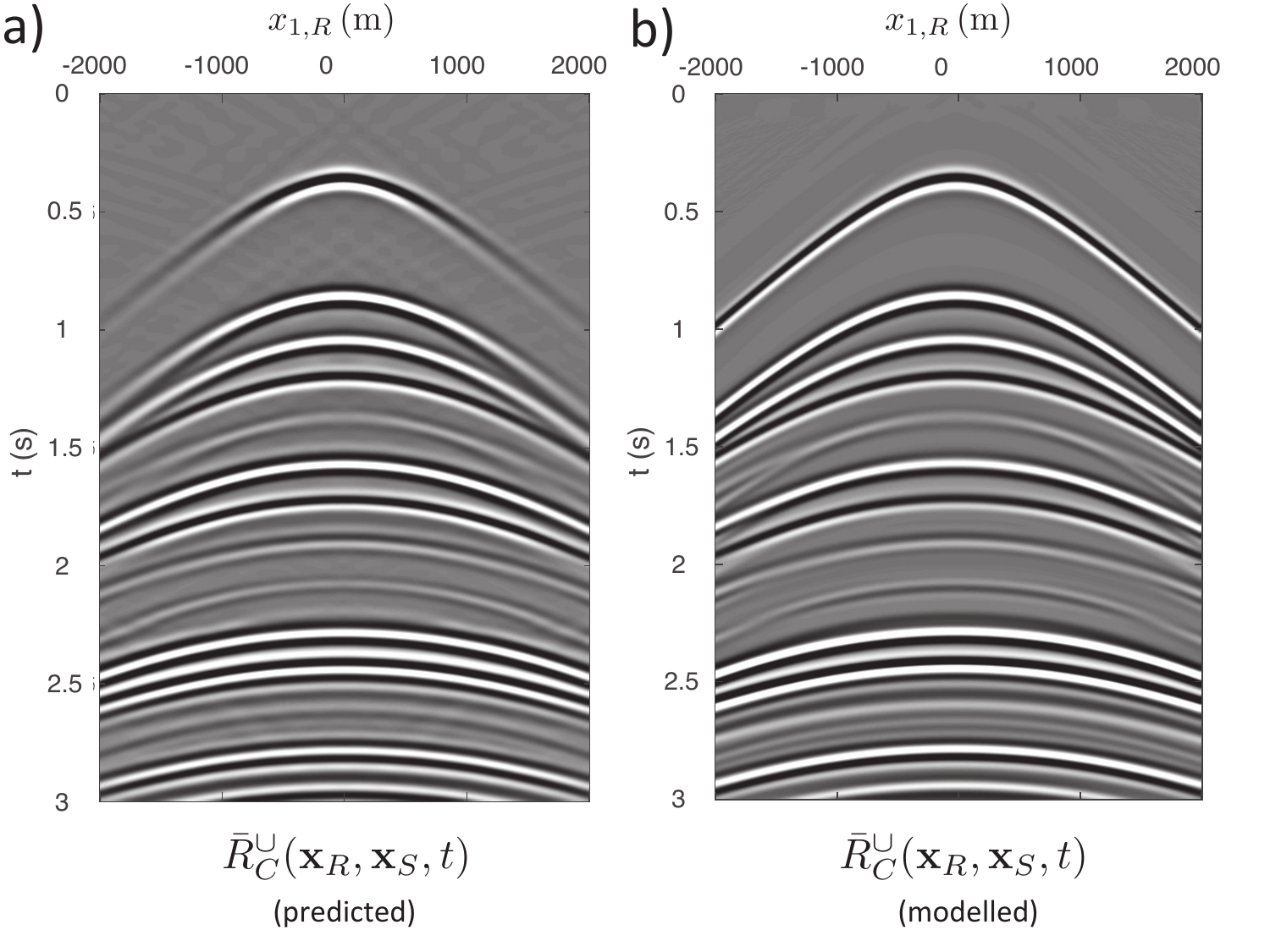}}
\vspace{-.2cm}
\caption{ 
(a) The predicted time-lapse response $\RC^\cup(\bx_R,\bx_S,t)$, constructed from \rev{$\RB^\cup(\bx_R,\bx_S,t)$ and the response of the underburden.}
(b) For comparison, the numerically modelled time-lapse response.}\label{Fig12}
\end{figure}

For the next example\rev{,} we consider a 2D acoustic point-source response of a horizontally layered medium. 
\rev{The medium is shown in Figure \ref{Fig11a}. Note that the overburden and underburden 
contain more layers than in the previous example.}
Figure \ref{Fig10}(a) shows the numerically modelled response $R_C^\cup(\bx_R,\bx_S,t)$ at the surface $\setdD_0$ in the time domain, for a fixed source at $\bx_S=(0,0)$ and variable receivers at $\bx_R=(x_{1,R},0)$.
Because the medium is horizontally layered, the responses to sources at other positions at $\setdD_0$ are simply laterally shifted versions of  the response in Figure \ref{Fig10}(a).
In the time-lapse scenario, the \rev{velocity} in the reservoir layer is changed from 3000 \rev{m/s} to 2500 \rev{m/s (and a similar change is applied to the mass density)}.
Figure \ref{Fig10}(b) shows the difference of the numerically modelled responses $R_C^\cup(\bx_R,\bx_S,t)$ and $\RC^\cup(\bx_R,\bx_S,t)$. 
\rev{The responses in this and the following figures are displayed with a small time-dependent gain of $\exp(0.5*t)$ to emphasise the internal multiples.}

\rev{We use our standard implementation of the Marchenko method \citep{Thorbecke2017GEO} for the estimation of the focusing functions.
Next, because the medium is horizontally layered, we efficiently carry out the layer replacement method 
in the wavenumber-frequency domain (hence, all integrals from equation (\ref{eq70kk}) onward reduce to straightforward products of the transformed quantities).} 
Figure \ref{Fig11}(a) shows the overburden response $R_A^\cup(\bx_R,\bx_S,t)$, resolved from equation (\ref{eq33kk}) \rev{in the wavenumber-frequency domain and transformed back to the space-time domain}. Note that the internal multiple\rev{s} of the \rev{overburden}, indicated by the arrow\rev{s},
\rev{have} been recovered from behind the  reflection \rev{response} of the reservoir layer.
The modelled response of the new target zone, $\Rb^\cup(\bx',\bx,t)$ at $\setdD_1$, is shown in Figure \ref{Fig11}(b),
for a fixed source at $\bx=(0,\rev{1400})$ m and variable receivers at $\bx'=(x_1',\rev{1400})$ m.
\rev{The predicted time-lapse reflection response at the surface of the overburden and target zone, $\RB^\cup(\bx_R,\bx_S,t)$, obtained with the representation 
of equation (\ref{eq4ag}) in the wavenumber-frequency domain,
is shown in  Figure \ref{Fig12a}(a). The numerically modelled time-lapse response  is shown (as a reference) in Figure \ref{Fig12a}(b).
Next, the response of the underburden is included, using the representation of equation (\ref{eq5ag}) in the wavenumber-frequency domain. This yields the 
 predicted time-lapse reflection response at the surface of the entire medium, $\RC^\cup(\bx_R,\bx_S,t)$, see Figure \ref{Fig12}(a). 
 The numerically modelled time-lapse response  of the entire medium is shown  in Figure \ref{Fig12}(b).}
Although the match is not as perfect as in the 1D example (Figure \ref{Fig9}(c)), Figure \ref{Fig12} shows that the 2D time-lapse response has been accurately
predicted. \rev{We used dip-filtering to suppress artefacts related to the finite aperture and the negligence of evanescent waves. This explains the diminishing 
amplitudes of the early reflections at large offsets.}

\newpage
\rev{\section{Discussion}\label{sec6}}

\rev{The numerical examples in the previous section 
show that under ideal circumstances the proposed method accurately predicts the time-lapse responses. Hence, these examples validate the theory. 
In practice there will be several factors that limit the accuracy. First, the direct arrivals of the focusing function $\f^\pm$,
needed to initiate 
the Marchenko scheme, are in practice defined in estimated models of the medium. Hence, the amplitudes and traveltimes of these direct arrivals will not be exact. 
The Marchenko method is robust to small-to-moderate errors in the direct arrival, in the sense that it predicts the multiples in the focusing functions and Green's functions, but these predicted 
multiples will exhibit similar amplitude and travel time errors as the direct arrival \citep{Wapenaar2014JASA, Broggini2014GEO}.
The errors in $\f^+$ and $\f^-$ largely compensate each other in the inversion of equation (\ref{eq33kk}), to obtain the
overburden response $\R_A^\cup$.
Hence, $\R_A^\cup$ will be retrieved very accurately, despite the errors in the direct arrival 
(it has been previously observed that the Marchenko method for obtaining data at the surface  is very robust \citep{Meles2016GEO, Neut2016GEO}).
This implies that multiples generated in the overburden are accurately separated from the response of deeper layers. 
The response of the overburden from below, $\R_A^\cap$, is obtained by inverting equation (\ref{eq333ll}). Here the amplitude errors in $\ff^+$ and $\ff^-$ largely compensate each other, but
travel time errors will result in an overall time shift of $\R_A^\cap$. A similar remark holds for the underburden response $\R_c^\cup$.
These errors will propagate into the predicted time-lapse response. We expect that the errors in the predicted primaries and low order multiples will be of the same order as the errors in the
direct arrivals and that these errors will grow for higher order multiples. 
}

\rev{The accuracy of the predicted time-lapse response will further be limited by losses in the medium, 
inaccuracies in the deconvolution for the source wavelet, the finite length of the acquisition aperture and incomplete sampling (particularly for 3D applications). 
Currently much research is going on to improve the Marchenko method to address these issues  \citep{Neut2016GEO, Ravasi2016GJI, Slob2016PRL, Staring2017SEG}. The proposed 
target replacement scheme will benefit from these developments.
}

\rev{The computational costs of the proposed method depend on the implementation. For the numerical examples in the previous section we took
advantage of the fact that the medium is horizontally layered. We implemented the 2D layer replacement in the wavenumber-frequency domain. This implies that
the inversion of the various integral equations is replaced by a straightforward scalar inversion per wavenumber-frequency combination. 
For laterally varying media, the integral equations should be solved in the space-frequency domain. After discretisation, this comes to a matrix inversion for each 
frequency component. In several cases (equations (\ref{eq99bb}) and (\ref{eq99bbcc})) 
the matrix inversion can efficiently be replaced by a series expansion, which can be terminated after a few terms, depending on the number of multiples that need 
to be taken into account. 
All at all, removing the target zone (section \ref{sec3}) requires applying the Marchenko method at two depth levels ($\setdD_1$ and $\setdD_2$)
and five matrix inversions (per frequency component) to solve integral equations (\ref{eq70kk}), (\ref{eq33kk}), 
(\ref{eq70ll}), (\ref{eq333ll}) and (\ref{eq3mm}). Inserting the new target zone (section \ref{sec4}) requires numerical modelling of the target zone response and 
 three matrix inversions (per frequency component) to solve integral equations (\ref{eq99bb}), (\ref{eq99bbcc}) and (\ref{eq101bc}).
 The costs for substituting the results into equations (\ref{eq4ag}) and (\ref{eq5ag}) are negligible in comparison with the matrix inversions.
Despite the significant number of steps for the entire process, the total costs should be seen in perspective with other methods. 
In comparison  with numerically modelling  the entire time-lapse reflection response, our method requires numerical modelling of the target zone response only.
The additional costs for  the Marchenko method and the matrix inversions are significant but not excessive. 
For example, applying the Marchenko method at two depth levels is feasible, considering the fact that 
some Marchenko imaging  methods apply this method for a large range of depth levels in an image volume \citep{Broggini2014GEO, Behura2014GEO}. 
The trade-off between the cost reduction for the numerical modelling and the cost increase related to the Marchenko method 
 and the matrix inversions  depends on the implementation details and  needs further investigation.
 }

\mbox{}\\
\section{Conclusions}\label{sec7}

We have proposed an efficient two-step process to replace the response of a target zone in a reflection response at the earth's surface.
In the  first step, the response of the original target zone is removed from the reflection response, using the Marchenko method. 
In the second step, the modelled response of a new target zone 
is inserted between the overburden and underburden responses. The method holds for vertically and laterally 
inhomogeneous lossless media. It fully accounts for all orders of multiple scattering and, in the elastodynamic case, for wave conversion.
It can be employed to predict the time-lapse reflection response for a range of target-zone scenarios. For this purpose, the first step needs to be carried out only once. 
Only the second step needs to be repeated  for each target-zone model. 
Since the target zone covers only a small part of the entire medium, repeated modelling of the target-zone response (and inserting it each time 
between the same overburden and underburden responses)  is a much more efficient process than repeated modelling of the  entire reflection 
response\rev{, but there are also additional costs related to the Marchenko method and several matrix inversions}.
This method may find applications in time-lapse full wave form inversion\rev{, for example to monitor 
fluid flow in an aquifer, subsurface storage of waste products, or production of a hydrocarbon reservoir}.
Since all multiples are taken into account, the coda following the  response of the target zone may be 
employed in the inversion. Because of the high sensitivity of the coda for changes in the medium \citep{Snieder2002Science}, this may ultimately
improve the resolution of the inverted time-lapse changes. 
\rev{Finally, when medium changes are not restricted to a reservoir, the target zone should be taken sufficiently large to include those parts of the embedding medium in which changes 
take place.} This will of course have a limiting effect on the efficiency gain. 

\acknowledgments

\rev{We thank Matteo Ravasi and an anonymous reviewer for their constructive remarks, which helped us to improve the paper. }
Figures \ref{Fig7} $-$ \ref{Fig9} have been generated with {\tt Matlab} scripts which \rev{can be found in the supporting information}. 
The synthetic data in Figures \ref{Fig10}, \ref{Fig11}(b),  \ref{Fig12a}(b) and \ref{Fig12}(b) have been modelled with Jan Thorbecke's finite difference code {\tt fdelmodc} 
(source code and manual can be found on {\tt https://janth.home.xs4all.nl}).
The research of KW has received funding from the European Research Council (ERC) under the European Union's Horizon 2020 research 
and innovation programme (grant agreement No: 742703).
The research of MS is part of the Dutch Open Technology Programme with project number 13939, which is financed by NWO Domain Applied and Engineering Sciences.

\appendix

\section*{\rev{Appendices}}

\section{Derivations for \rev{S}ection \ref{sec3}}

\subsection{Representations for Marchenko method}\label{AppD}

\begin{table}
\caption{Quantities to derive Marchenko representations.}\label{tableA1}
\begin{center}
\begin{tabular}{ccc}
\hline
& State $A$: & State $B$: \\
&Medium $C$&Medium $A$\\
&Source at $\bxr$ just above $\setdD_0$&Focus at $\bx'$ at $\setdD_1$\\
\hline
 $\setdD_0$   & $\pa^+(\bx,\omega)\too\I\delta(\bxh-\bxhr)$ & $\pb^+(\bx,\omega)\too\f^+(\bx,\bx',\omega)$\\
 &$\hspace{1.2cm}+\r \R_C^\cup(\bx,\bxr,\omega)$&$\hspace{1.2cm}+\r\f^-(\bx,\bx',\omega)$\\
   & $\pa^-(\bx,\omega)\too\R_C^\cup(\bx,\bxr,\omega)$ & $\pb^-(\bx,\omega)\too\f^-(\bx,\bx',\omega)$\\
\hline
 $\setdD_1$   &  $\pa^+(\bx,\omega)\too\G_C^{+,+}(\bx,\bxr,\omega)$ & $\pb^+(\bx,\omega)\too\I\delta(\bxh-\bxh')$\\
  & $\pa^-(\bx,\omega)\too\G_C^{-,+}(\bx,\bxr,\omega)$ & $\pb^-(\bx,\omega)\too\O$ \\
\hline
\end{tabular}
\end{center}
\end{table}
We derive relations between  
$\R_C^\cup$,  $\f^\pm$  and $\G_C^{-,\pm}$.
State $A$ in Table A.1 is defined in a similar way as state $B$ in Table 1, except that here we consider medium $C$, and we choose a source at $\bxr$, just above $\setdD_0$.
State $B$ in Table A.1 represents the focusing function, which is defined in medium $A$. 
At $\setdD_0$, the downgoing field consists of the emitted focusing function $\f^+(\bx,\bx',\omega)$,
plus the downward reflected upgoing part of the focusing function. The latter term is absent when the earth's surface is transparent.
The upgoing field at $\setdD_0$ is given by the upgoing part of the focusing function.
The quantities at $\setdD_1$  in state $B$ represent the focusing conditions, formulated by equations (\ref{eq25}) and (\ref{eq26}).

We substitute the quantities of Table A.1 into equation (\ref{eq1}).
Using equations 
(\ref{eqex6})  and (\ref{eqex11}), setting $m=0$ and $n=1$, this gives
\begin{eqnarray}\label{eq72}
\hspace{.0cm}&&\{\G_C^{-,+}(\bx',\bxr,\omega)\}^t+\f^-(\bxr,\bx',\omega)
=\int_{\setdD_0}\R_C^\cup(\bxr,\bx,\omega)\f^+(\bx,\bx',\omega){\rm d}\bx,
\end{eqnarray}
for $\bxr$ just above $\setdD_0$ and $\bx'$ at $\setdD_1$.
Next, we substitute the quantities of Table A.1 into equation (\ref{eq1b}).
Using equations 
(\ref{eqex6}) and  (\ref{eqex11}), setting $m=0$ and $n=1$, this gives
\begin{eqnarray}\label{eq73}
&&\hspace{-0.7cm}\{\G_C^{+,+}(\bx',\bxr,\omega)\}^t-\{\f^+(\bxr,\bx',\omega)+\r\f^-(\bxr,\bx',\omega)\}^*\nonumber\\
&&\hspace{-0.7cm}=\int_{\setdD_0}\R_C^\cup(\bxr,\bx,\omega)\r\{\f^+(\bx,\bx',\omega)\}^*{\rm d}\bx
\nonumber\\&&\hspace{-0.7cm}
-\int_{\setdD_0}\R_C^\cup(\bxr,\bx,\omega)\{\I-(\r)^\dagger\r\}^*\{\f^-(\bx,\bx',\omega)\}^*{\rm d}\bx,
\end{eqnarray}
for $\bxr$ just above $\setdD_0$ and $\bx'$ at $\setdD_1$.
Equations (\ref{eq72}) and (\ref{eq73}) hold for the situation with or without free surface just above $\setdD_0$. Equation (\ref{eq73}) can be further simplified for each of these situations.
For the situation without free surface, with $\r=\O$, equation (\ref{eq73}) becomes
\begin{eqnarray}\label{eq73a}
\hspace{.0cm}&&\{\G_C^{+,+}(\bx',\bxr,\omega)\}^t-\{\f^+(\bxr,\bx',\omega)\}^*
=-\int_{\setdD_0}\R_C^\cup(\bxr,\bx,\omega)\{\f^-(\bx,\bx',\omega)\}^*{\rm d}\bx.
\end{eqnarray}
On the other hand, for the situation with free surface, with $(\r)^\dagger\r=\I$ (equation (\ref{eqex12})), we obtain
\begin{eqnarray}\label{eq73b}
&&\hspace{-0.7cm}\{\G_C^{+,+}(\bx',\bxr,\omega)\}^t-\{\f^+(\bxr,\bx',\omega)+\r\f^-(\bxr,\bx',\omega)\}^*
\nonumber\\&&\hspace{-0.7cm}
=\int_{\setdD_0}\R_C^\cup(\bxr,\bx,\omega)\r\{\f^+(\bx,\bx',\omega)\}^*{\rm d}\bx.
\end{eqnarray}

\subsection{Response to the focusing function $\f^+$}\label{AppA}

\begin{table}
\caption{Quantities to derive the response to $\f^+$.}\label{tableA2}
\begin{center}
\begin{tabular}{ccc}
\hline
& State $A$: & State $B$: \\
&Medium $A$&Medium $A$\\
&Source at $\bx''$ just below $\setdD_1$&Focus at $\bx'$ at $\setdD_1$\\
\hline
 $\setdD_0$  & $\pa^+(\bx,\omega)\too\r\T_A^-(\bx,\bx'',\omega)$ & $\pb^+(\bx,\omega)\too\f^+(\bx,\bx',\omega)$\\
  &&$\hspace{1.2cm}+\r\f^-(\bx,\bx',\omega)$\\
 & $\pa^-(\bx,\omega)\too\T_A^-(\bx,\bx'',\omega)$ & $\pb^-(\bx,\omega)\too\f^-(\bx,\bx',\omega)$\\
\hline
 $\setdD_1$  & $\pa^+(\bx,\omega)\too\R_A^\cap(\bx,\bx'',\omega)$ &  $\pb^+(\bx,\omega)\too\I\delta(\bxh-\bxh')$\\
  & $\pa^-(\bx,\omega)\too\I\delta(\bxh-\bxh'')$ & $\pb^-(\bx,\omega)\too\O$\\
\hline
\end{tabular}
\end{center}
\end{table}

We derive the response to the focusing function $\f^+(\bx,\bx',\omega)$, when emitted into medium $A$ from above.
For state $A$ in Table A.2 we place a source in medium $A$ at $\bx''$, just below $\setdD_1$. The flux-normalised upgoing field at $\setdD_1$ is the delta function $\I\delta(\bxh-\bxh'')$,
with its a singularity vertically above the source. There are no other contributions to this upgoing field because the medium below $\setdD_1$ is homogeneous.
The downgoing field at $\setdD_1$ is the reflection response of medium $A$ from below, $\R_A^\cap(\bx,\bx'',\omega)$. At $\setdD_0$, the upgoing field is the transmission response
$\T_A^-(\bx,\bx'',\omega)$ and the downgoing field is given by the downward reflected transmission response. The latter vanishes when the earth's surface is transparent.
For state $B$ we choose the same focusing function as in Table A.1. 
We substitute the quantities of Table A.2 into equation (\ref{eq1}).
Using equations (\ref{eqex11}) and (\ref{eqex14}), setting $m=0$ and $n=1$, this gives
\begin{equation}\label{eq70}
\I\delta(\bxh''-\bxh')=\int_{\setdD_0}\T_A^+(\bx'',\bx,\omega)\f^+(\bx,\bx',\omega){\rm d}\bx,
\end{equation}
for $\bx'$  at $\setdD_1$ and  $\bx''$ just below $\setdD_1$.
Since $\setdD_1$ is transparent, $\bx''$ may just as well be chosen at $\setdD_1$.

To derive the reflection response to the focusing function $\f^+$, we combine state $A$ of Table 1 with state $B$ of Table A.2. 
Substitution of these  quantities into equation (\ref{eq1}), 
using equations (\ref{eqex6}) and (\ref{eqex11}), setting $m=0$ and $n=1$,  gives
\begin{equation}\label{eq33}
\f^-(\bxr,\bx',\omega)=\int_{\setdD_0}\R_A^\cup(\bxr,\bx,\omega)\f^+(\bx,\bx',\omega){\rm d}\bx,
\end{equation}
for $\bxr$ just above $\setdD_0$ and $\bx'$ at $\setdD_1$.

\subsection{Response to the focusing function $\ff^-$}\label{AppB}

\begin{table}
\caption{Quantities to derive the response to $\ff^-$.}\label{tableA3}
\begin{center}
\begin{tabular}{ccc}
\hline
& State $A$: & State $B$: \\
&Medium $A$&Medium $A$\\
&Source at $\bx''$ just above $\setdD_0$&Focus at $\bx'$ at $\setdD_0$\\
\hline
 $\setdD_0$  & $\pa^+(\bx,\omega)\too\I\delta(\bxh-\bxh'')$ & $\pb^+(\bx,\omega)\too\r\I\delta(\bxh-\bxh')$\\
&\hspace{1.2cm}$+\r\R_A^\cup(\bx,\bx'',\omega)$&\\
  & $\pa^-(\bx,\omega)\too\R_A^\cup(\bx,\bx'',\omega)$ & $\pb^-(\bx,\omega)\too\I\delta(\bxh-\bxh')$\\
\hline
 $\setdD_1$  & $\pa^+(\bx,\omega)\too\T_A^+(\bx,\bx'',\omega)$ &  $\pb^+(\bx,\omega)\too\ff^+(\bx,\bx',\omega)$\\
  & $\pa^-(\bx,\omega)\too\O$ & $\pb^-(\bx,\omega)\too\ff^-(\bx,\bx',\omega)$\\
\hline
\end{tabular}
\end{center}
\end{table}
We derive the response to the focusing function $\ff^-(\bx,\bx',\omega)$, when emitted into medium $A$ from below.
For state $A$ in Table A.3 we place a source in medium $A$ at $\bx''$, just above $\setdD_0$. This needs no further explanation, because this is very similar to state $A$ in Table 1.
State $B$ represents the focusing function, which is defined in medium $A$. At $\setdD_1$, the upgoing field is given by the emitted focusing function $\ff^-(\bx,\bx',\omega)$.
There are no other contributions to this upgoing field because the medium below $\setdD_1$ is homogeneous.
The downgoing field at $\setdD_1$ is given by the downgoing part of the focusing function.
The quantities at $\setdD_0$  in state $B$ represent the focusing conditions, formulated by equations (\ref{eq25a}) and (\ref{eq26a}).

We substitute the quantities of Table A.3 into equation (\ref{eq1}).
Using equations (\ref{eqex11}) and (\ref{eqex14}), setting $m=0$ and $n=1$, this gives
\begin{equation}\label{eq70g}
\I\delta(\bxh''-\bxh')=\int_{\setdD_1}\T_A^-(\bx'',\bx,\omega)\ff^-(\bx,\bx',\omega){\rm d}\bx,
\end{equation}
for $\bx'$  at $\setdD_0$ and  $\bx''$ just above $\setdD_0$.
Since $\setdD_0$ is transparent, $\bx''$ may just as well be chosen at $\setdD_0$.

To derive the reflection response to the focusing function $\ff^-$, we combine state $A$ of Table A.2 with state $B$ of Table A.3. 
Substitution of these  quantities into equation (\ref{eq1}), 
using equations (\ref{eqex7}) and (\ref{eqex11}), setting $m=0$ and $n=1$, gives
\begin{equation}\label{eq333}
\ff^+(\bx'',\bx',\omega)=\int_{\setdD_1}\R_A^\cap(\bx'',\bx,\omega)\ff^-(\bx,\bx',\omega){\rm d}\bx,
\end{equation}
for $\bx'$ at $\setdD_0$ and $\bx''$ just below $\setdD_1$.

\subsection{Relations between  $\f^\pm$ and $\ff^\pm$}\label{AppC}

To derive the relations between $\f^\pm$ and $\ff^\pm$, we take for state $A$ the quantities defined in Table A.3 for state $B$ and replace $\bx'$ by $\bx''$.
For state $B$ we take the quantities defined in Table A.2 for state $B$.
Substitution of these  quantities into equation (\ref{eq1}), 
using equation (\ref{eqex11}), setting $m=0$ and $n=1$, gives
\begin{equation}\label{eq55}
\f^+(\bx'',\bx',\omega)=\{\ff^-(\bx',\bx'',\omega)\}^t,
\end{equation}
for $\bx''$ at $\setdD_0$ and $\bx'$ at $\setdD_1$.
Substituting the same quantities into equation (\ref{eq1b}), using equation (\ref{eqex11}), setting $m=0$ and $n=1$, gives
\begin{eqnarray}\label{eq56}
&&\{\I-(\r)^\dagger\r\}\f^-(\bx'',\bx',\omega)-(\r)^*\f^+(\bx'',\bx',\omega)
=-\{\ff^+(\bx',\bx'',\omega)\}^\dagger.
\end{eqnarray}
Equations (\ref{eq55}) and (\ref{eq56}) hold for the situation with or without free surface just above $\setdD_0$. Equation (\ref{eq56}) can be further simplified for each of these situations.
For the situation without free surface, with $\r=\O$, equation (\ref{eq56}) becomes
\begin{equation}\label{eq56a}
\f^-(\bx'',\bx',\omega)=-\{\ff^+(\bx',\bx'',\omega)\}^\dagger.
\end{equation} 
On the other hand, for the situation with free surface, with $(\r)^\dagger\r=\I$ (equation (\ref{eqex12})), we obtain
\begin{equation}\label{eq56b}
(\r)^*\f^+(\bx'',\bx',\omega)=\{\ff^+(\bx',\bx'',\omega)\}^\dagger.
\end{equation} 
Using equation (\ref{eq55}) this gives the following symmetry relation for $\ff^\pm$
\begin{equation}\label{eq56c}
(\r)^*\{\ff^-(\bx',\bx'',\omega)\}^t=\{\ff^+(\bx',\bx'',\omega)\}^\dagger.
\end{equation} 

\section{Derivations for \rev{S}ection \ref{sec4}}

\subsection{Equation for $\GBb^{+,+}(\bx,\bxs,\omega)$}\label{AppE}

To derive an equation for $\GBb^{+,+}(\bx,\bxs,\omega)$, we take for state $A$ the quantities defined in Table A.2 for state $A$.
For state $B$ we take the quantities defined in Table 1 for state $B$, \rev{but with bars on these quantities}.
Substitution of these quantities into equation (\ref{eq1}),  using equations (\ref{eqex7}), (\ref{eqex11}) and  (\ref{eqex14}), setting $m=0$ and $n=1$, gives
\begin{eqnarray}\label{eqB1}
\hspace{.2cm}\T_A^+(\bx'',\bxs,\omega)&=&\GBb^{+,+}(\bx'',\bxs,\omega)
-\int_{\setdD_1}\R_A^\cap(\bx'',\bx,\omega)\GBb^{-,+}(\bx,\bxs,\omega){\rm d}\bx, 
\end{eqnarray}
for $\bxs$ just above $\setdD_0$ and $\bx''$ just below $\setdD_1$.
Since $\setdD_1$ is transparent, $\bx''$ may just as well be chosen at $\setdD_1$.
Next, we replace the integration variable $\bx$ by $\bx'$ and substitute equation (\ref{eq3}) \rev{(but with bars on all quantities)}
into the right-hand side of equation (\ref{eqB1}). This gives
\begin{eqnarray}\label{eq8}
&&\hspace{-1.3cm}\T_A^+(\bx'',\bxs,\omega)=\GBb^{+,+}(\bx'',\bxs,\omega)
-\int_{\setdD_1}\int_{\setdD_1}\R_A^\cap(\bx'',\bx',\omega)\Rbb^\cup(\bx',\bx,\omega)\GBb^{+,+}(\bx,\bxs,\omega){\rm d}\bx{\rm d}\bx',
\end{eqnarray}
for $\bxs$ just above $\setdD_0$ and $\bx''$ at $\setdD_1$.
We can rewrite this as
\begin{equation}\label{eq99}
\T_A^+(\bx'',\bxs,\omega)=\int_{\setdD_1}\MAbb(\bx'',\bx,\omega)\GBb^{+,+}(\bx,\bxs,\omega){\rm d}\bx,
\end{equation}
with
\begin{eqnarray}\label{eq999}
\hspace{.5cm}\MAbb(\bx'',\bx,\omega)&=&\I\delta(\bxh''-\bxh)
-\int_{\setdD_1}\R_A^\cap(\bx'',\bx',\omega)\Rbb^\cup(\bx',\bx,\omega){\rm d}\bx', 
\end{eqnarray}
for $\bx$ and  $\bx''$ at $\setdD_1$.

\subsection{Representation for $\TBb^+(\bx'',\bxs,\omega)$}\label{AppF}

\begin{table}
\caption{Quantities to derive representation for $\TBb^+(\bx'',\bxs,\omega)$.}\label{tableB1}
\begin{center}
\begin{tabular}{ccc}
\hline
& State $A$: & State $B$: \\
&Medium $\bar b$&Medium $\bar B$\\
&Source at $\bx''$ just below $\setdD_2$&Source at $\bxs$ just above $\setdD_0$\\
\hline
 $\setdD_1$  & $\pa^+(\bx,\omega)\too\O$ & $\pb^+(\bx,\omega)\too\GBb^{+,+}(\bx,\bxs,\omega)$\\
  & $\pa^-(\bx,\omega)\too\Tbb^-(\bx,\bx'',\omega)$ & $\pb^-(\bx,\omega)\too\GBb^{-,+}(\bx,\bxs,\omega)$\\
\hline
 $\setdD_2$  & $\pa^+(\bx,\omega)\too\Rbb^\cap(\bx,\bx'',\omega)$ &  $\pb^+(\bx,\omega)\too\TBb^+(\bx,\bxs,\omega)$\\
  & $\pa^-(\bx,\omega)\too\I\delta(\bxh-\bxh'')$ & $\pb^-(\bx,\omega)\too\O$\\
\hline
\end{tabular}
\end{center}
\end{table}

We derive a representation for $\TBb^+(\bx'',\bxs,\omega)$, in terms of the Green's function $\GBb^{+,+}(\bx,\bxs,\omega)$ and the transmission response of unit $\bar b$, 
$\Tbb^+(\bx'',\bx,\omega)$. 
Substituting the quantities of Table B.1 into equation (\ref{eq1}), using 
equation (\ref{eqex14}), setting $m=1$ and $n=2$, gives
\begin{equation}\label{eq15m}
\TBb^+(\bx'',\bxs,\omega)=\int_{\setdD_1}\Tbb^+(\bx'',\bx,\omega)\GBb^{+,+}(\bx,\bxs,\omega){\rm d}\bx,
\end{equation}
for $\bxs$ just above $\setdD_0$ and  $\bx''$ just below $\setdD_2$.
Since $\setdD_2$ is transparent, $\bx''$ may just as well be chosen at $\setdD_2$.

\subsection{Equation for $\RBb^\cap(\bx,\bx',\omega)$}\label{AppG}

\begin{table}
\caption{Quantities to derive equation for $\RBb^\cap(\bx,\bx',\omega)$.}\label{tableB2}
\begin{center}
\begin{tabular}{ccc}
\hline
& State $A$: & State $B$: \\
&Medium $\bar B$&Medium $\bar B$\\
&Source at $\bxs$ just above $\setdD_0$&Source at $\bx'$ just below $\setdD_2$\\
\hline
 $\setdD_0$  & $\pa^+(\bx,\omega)\too\I\delta(\bxh-\bxhs)$ & $\pb^+(\bx,\omega)\too\r\TBb^-(\bx,\bx',\omega)$\\
 &$\hspace{1.2cm}+\r\RBb^\cup(\bx,\bxs,\omega)$&\\
  & $\pa^-(\bx,\omega)\too\RBb^\cup(\bx,\bxs,\omega)$ & $\pb^-(\bx,\omega)\too\TBb^-(\bx,\bx',\omega)$\\
\hline
 $\setdD_2$  & $\pa^+(\bx,\omega)\too\TBb^+(\bx,\bxs,\omega)$ &  $\pb^+(\bx,\omega)\too\RBb^\cap(\bx,\bx',\omega)$\\
  & $\pa^-(\bx,\omega)\too\O$ & $\pb^-(\bx,\omega)\too\I\delta(\bxh-\bxh')$\\
\hline
\end{tabular}
\end{center}
\end{table}
We derive an equation for $\RBb^\cap(\bx,\bx',\omega)$.
Substituting the quantities of Table B.2 into equation (\ref{eq1b}), using equations (\ref{eqex6}) and (\ref{eqex14}),
setting $m=0$ and $n=2$, gives
\begin{eqnarray}
&&\hspace{-.5cm}\int_{\setdD_2}\{\TBb^-(\bxs,\bx,\omega)\}^*\RBb^\cap(\bx,\bx',\omega){\rm d}\bx=\r\TBb^-(\bxs,\bx',\omega)\nonumber\\
&&\hspace{-.5cm}-\int_{\setdD_0}\{\RBb^\cup(\bxs,\bx,\omega)\}^*\{\I-(\r)^\dagger\r\}\TBb^-(\bx,\bx',\omega){\rm d}\bx,\label{eq101}
\end{eqnarray}
for $\bxs$ just above $\setdD_0$ and  $\bx'$ just below $\setdD_2$.
Since $\setdD_2$ is transparent, $\bx'$ may just as well be chosen at $\setdD_2$.
For the situation without free surface, with $\r=\O$, this gives
\begin{eqnarray}
\hspace{.7cm}&&\int_{\setdD_2}\{\TBb^-(\bxs,\bx,\omega)\}^*\RBb^\cap(\bx,\bx',\omega){\rm d}\bx=
-\int_{\setdD_0}\{\RBb^\cup(\bxs,\bx,\omega)\}^*\TBb^-(\bx,\bx',\omega){\rm d}\bx.\label{eq101b}
\end{eqnarray}
On the other hand, for the situation with free surface, with $(\r)^\dagger\r=\I$ (equation \ref{eqex12}), we obtain
\begin{equation}
\int_{\setdD_2}\{\TBb^-(\bxs,\bx,\omega)\}^*\RBb^\cap(\bx,\bx',\omega){\rm d}\bx=\r\TBb^-(\bxs,\bx',\omega).
\end{equation}

\end{document}